%% file: parameter_extraction_JES_Final.tex
\documentclass[final,3p,times]{elsarticle}

\input{headinput}

\usepackage{amssymb}
\usepackage[english]{babel}
\usepackage[utf8x]{inputenc}
\usepackage{amsmath}
\DeclareMathOperator{\tr}{tr}
\usepackage[overload]{empheq}
\usepackage{framed}
\usepackage{marginnote}
\usepackage[toc,page]{appendix}
\usepackage{textcomp}

\allowdisplaybreaks

\biboptions{numbers,sort&compress}

\usepackage{lipsum}

\usepackage{todonotes}
\usepackage{multirow,booktabs}
\usepackage{graphicx}
\usepackage{amsfonts}
\graphicspath{{images/}}
\usepackage{subfigure}
\usepackage[justification=centering]{caption}
\usepackage{epstopdf}
\usepackage{gensymb}
\usepackage[flushleft]{threeparttable}
\usepackage{balance} 
\usepackage{bm}

\usepackage{float}

\usepackage{algorithmic}
\usepackage{algorithm}

\allowdisplaybreaks

\newtheorem{definition}{Definition}

\usepackage{setspace}

 \makeatletter
\def\ps@pprintTitle{%
 \let\@oddhead\@empty
 \let\@evenhead\@empty
 \def\@oddfoot{\centerline{\thepage}}%
 \let\@evenfoot\@oddfoot}
\makeatother

\begin{document}

\begin{frontmatter}




\title{\LARGE {\bf
One-Shot Parameter Identification of the Thevenin's Model for Batteries: Methods and Validation\tnoteref{Sponsorship}}
}

\tnotetext[Sponsorship]{This work was supported in part by the United States National Science Foundation   under Awards CMMI-1763093 and CMMI-1847651.}

\author[label1]{Ning~Tian}\author[label2]{Yebin~Wang}\author[label3]{Jian~Chen}\author[label1]{Huazhen~Fang\corref{cor1}}
\address[label1]{Department of Mechanical Engineering, University of Kansas, Lawrence, KS 66045, USA}
\address[label2]{Mitsubishi Electric Research Laboratories, Cambridge, MA 02139, USA}
\address[label3]{State Key Laboratory of Industrial Control Technology,  College of Control Science \& Engineering, Zhejiang University, Hangzhou 310027, China}
 

 \cortext[cor1]{Corresponding author. E-mail address: fang@ku.edu.}
 


\date{}

\begin{abstract}
Parameter estimation is of foundational importance for various model-based battery management tasks, including  charging control, state-of-charge estimation and aging assessment. However, it remains a challenging issue as the existing methods generally depend on cumbersome and  time-consuming  procedures to extract battery parameters from data. 
 Departing from the literature, this paper sets the unique aim of identifying all the parameters offline in a one-shot procedure, including the resistance and capacitance parameters and the parameters in the parameterized function mapping from the state-of-charge to the open-circuit voltage. Considering the well-known Thevenin's battery model, the study begins with the parameter identifiability analysis, showing that all the parameters are locally identifiable. Then, it formulates the parameter identification problem  in a prediction-error-minimization framework. As  the non-convexity  intrinsic to the problem  may lead to physically meaningless estimates,  two  methods are  developed to overcome this issue. The first one is to constrain the parameter search within a reasonable space by setting parameter bounds, and the other adopts regularization of the cost function using prior parameter guess.  The proposed identifiability analysis and identification methods are extensively validated through simulations and experiments. 
\end{abstract}


\end{frontmatter}

\section{Introduction}\label{Sec:Introduction}
Rechargeable batteries have found wide use nowadays in the  consumer electronics, transportation and grid sectors by the billions. Their ever-widening use has excited an intense interest  in advanced battery management system research. The main subjects of inquiry in this area include state-of-charge (SoC) and state-of-health (SoH) estimation, optimal charging control, cell balancing, thermal management, e.g.,~\citep{coleman2007state,lee2008state, Perez:ACC:2015,Wang:CS:2017, lin2003power,Fang:TCST:2017,Fang:JES:2018,Perez:TVT:2017,Liu:TCST:2017, Ouyang:TSE:2018,lin2014lumped,Tian:TCST:2017}, and the references therein.
Playing a foundational role in many of the existing studies are the equivalent circuit models (ECMs), which replicate a battery's electrical dynamics using a circuit composed of resistors, capacitors and voltage sources. However, the parameters of an ECM are often   unknown in many real-world circumstances or drift with  cycled charging/discharging. This   presents the question of how to accurately and efficiently identify them from the current/voltage measurements made on a battery cell.  

The current literature on ECM identification can be roughly divided into three categories:
\begin{itemize}
\item {\bf Experiment-based analysis.} This category develops and conducts charging/discharging experiments such that a battery's dynamics  can be   exposed and used  to estimate parameters. For instance, the transient voltage responses are leveraged in~\citep{dubarry2007development,schweighofer2003modeling}
 to identify the internal resistance and RC-based time constants by charging or discharging a battery using constant or pulse currents. Another example is the experimental determination of the relationship between SoC and open-circuit voltage (OCV). Conventionally, it can be accomplished by charging or discharging the battery using a very small current~\citep{plett2004extended,weng2014unified,xing2014state}, or alternatively,  applying a current of normal magnitude intermittently   (a sufficiently long rest period is applied between two discharging operations)~\citep{he2011state,tian2014modified,petzl2013advancements}. While easy to implement, these approaches introduce significant time costs---an SoC-OCV calibration experiment can take more than one day~\citep{plett2004extended,
dubarry2007development},  unaffordable especially in massive battery testing.  One can also find studies about the design of specialized charging/discharging protocols to expedite parameter identification, e.g.~\cite{abu2004rapid}, which, however, would still take more than ten hours.

\item{\bf Electrochemical impedance spectroscopy (EIS).} EIS is an important means of observing electrochemical processes within batteries. The EIS data  reveal the impedance properties of a battery, and the literature includes a few methods that fit an ECM to  collected EIS data to extract the resistance and RC parameters~ \citep{buller2003impedance,nelatury2004equivalent,
goebel2008prognostics,andre2011characterization}. These methods focus on  impedance identification as needed in many applications and meanwhile, leave  other parts of a battery's dynamics such as the SoC-OCV function  beyond consideration.

\item {\bf Analytical data-based parameter estimation.}  This category seeks to determine an ECM's unknown parameters using the current/voltage measurement  data from a system identification perspective. The studies in~\citep{chen2006accurate,yang2016improved,baronti2010enhanced,lam2011practical,
hentunen2014time,gandolfo2015dynamic,zheng2009dynamic,kim2011hybrid} consider pulse current charging/discharging experiments and use the data to identify an ECM's RC parameters  by fitting the model to the pulse phase voltage~\citep{chen2006accurate,
yang2016improved}, the relaxation phase voltage~\citep{baronti2010enhanced,
lam2011practical,hentunen2014time,gandolfo2015dynamic}, or both of them~\cite{zheng2009dynamic,kim2011hybrid}. Some other studies perform similar data-fitting-based RC parameter estimation using data from variable current charging/discharging tests, while exploiting generic optimization~\cite{einhorn2012comparison}, grey-box optimization~\cite{birkl2013model}, sequential quadratic programming~\cite{li2012new}, or particle swarm optimization~\citep{hu2012comparative,yu2017model} as the solution tools. The studies in~\citep{gu1983mathematical,Hu:JPS:2013,devarakonda2014algebraic,hu2013model} propose to estimate the RC parameters by converting the identification problem into a problem of solving a set of linear and polynomial equations.  In~\citep{Hu:JPS:2011,Li:IJER:2014}, a linear state-space model is formulated for batteries, and subspace identification is then performed to infer the system matrices. An approach presented in~\citep{xia2016accurate} directly identifies the parameters of a continuous-time ECM using sampled discrete-time measurements. Associated with parameter identification, there is a growing amount of work on combined  estimation of SoC and part of the model parameters~\citep{chiang2011online,Sitterly:TSE:2011,tang2011li,fang2014state,
he2012online,rahimi2014online,fleischer2014line,feng2015online,
Ye:Energy:2018,Zhang:Energy:2018}. It is noted that these studies usually require an accurate SoC-OCV relationship to be available prior to identification, which requires long-time testing as aforementioned. 

\end{itemize}

Despite the importance, the methods surveyed above, however, share one limitation: they are designed to identify only a subset of an ECM's parameters, on the premise  that the other parameters are known. This 
brings about an intriguing question: {\em Is it  possible to extract   the parameters of an ECM all at once?} Here, by ``all'', it means the RC parameters as well as the parameters of the nonlinear SoC-OCV function.  At least two benefits can result if this can be achieved. First, it will enhance the efficiency of battery model identification considerably by avoiding the tedious SoC-OCV calibration. Second, it can help ensure the availability of an accurate model for   battery management   during a battery's service life. However, one-shot identification for batteries has been long known as a challenge, because of  the increase in the number of parameters to estimate and the serious nonlinearity and non-convexity  in the identification procedure. The only studies on this topic, to our knowledge, are given in~\citep{hu2009technique,hu2011electro,brand2014extraction,malik2014extraction}, which  use the genetic algorithms to search for the best  parameter estimates for a battery model. However, their application  still face some limitations. The primary one among them is the weak parameter identifiability   due to the many parameters (more than 30)   to   determine, which can potentially compromise the estimation accuracy. In addition, the genetic algorithms  generally converge slowly and impose high computational expenses.

This work is motivated to develop new and efficient approaches for one-shot battery parameter identification.  The Thevenin's model, which has been a popular choice for battery  management~\citep{rao2003battery,jongerden2008battery,
Plett:Artech:2015,seaman2014survey}, is considered here. 
This model is nonlinear by nature, and consequently, an identification effort can suffer pitfalls caused by the  nonlinearity and non-convexity, which may eventually produce inaccurate or unphysical parameter estimation. 
This  work hence  presents a systematic study to overcome such an issue, yielding the following contributions.
\begin{itemize}
\item The study  for the first time reveals the feasibility of the one-shot identification for the Thevenin's model through an in-depth  parameter identifiability analysis. Specifically, it shows that all the parameters are locally identifiable, indicating that they can be uniquely determined in a local domain. 

\item With the model's local identifiability guaranteed, this work   synthesizes novel one-shot parameter identification methods. The methods are developed to minimize the model prediction error through numerical optimization. Different from the literature,  we introduce two critical mechanisms     to address  the non-convexity issue: 1) constrained optimization, which    constrains the search space by applying upper and lower bounds to some parameters, and  2) generalized Tikhonov regularization, which adds a regularization term to the considered cost function to drive the optimization toward a reasonable minimum point. 

\item  The theoretical estimation accuracy of  the proposed identification methods  is rigorously characterized. The approaches and the associated   analysis are   validated using extensive simulations and experiments. 
\end{itemize}

With the above contributions, our study can   enable easier availability of   accurate Thevenin's models for advanced battery management ranging from  charging control to SoC estimation and aging prognostics.

 The rest of the paper is organized as follows.  Section~\ref{Sec:Problem-Formulation} reviews the Thevenin's model and sets up the parameter identification problem. Section~\ref{Sec:Parameter-Identification} investigates the parameter identifiability and develops the  identification approaches.  Section~\ref{Sec:Numerical-Simulation} evaluates the efficacy of the proposed results through Monte Carlo simulations. Section~\ref{Sec:Experimental-Validation} further presents validation based on experiments. Finally, some concluding remarks are gathered in Section~\ref{Sec:Conclusion}.

\section{Thevenin's Model}\label{Sec:Problem-Formulation}
This section first introduces the Thevenin's model and then derives the voltage response   under constant-current discharging. The formulation of  the parameter identification problem then follows. 

\begin{figure}[h]
\centering
\includegraphics[trim = {56mm 75mm 70mm 51mm}, clip, width=0.47\textwidth]{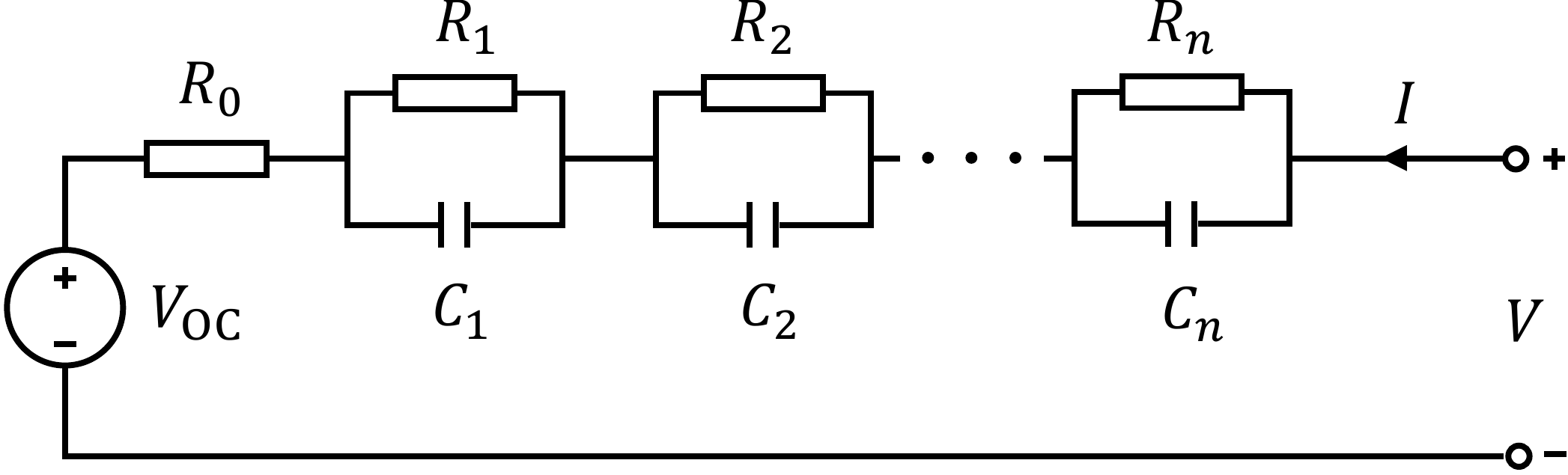}
\centering
\caption{The Thevenin's equivalent circuit model.}
\label{Fig:Thevenin-Model}
\vspace{-5mm}
\end{figure}

A general form of the Thevenin's model is shown in Figure~\ref{Fig:Thevenin-Model}. The first main component of the model is a voltage source, which emulates a battery's OCV. 
The OCV is SoC-dependent, and
the SoC's dynamics is given by
\begin{align}\label{SoC-Counting}
\dot{\mathrm{SoC}} & = \frac{1}{3600Q_{\rm c}}I ,
\end{align}
where $Q_{\rm c}$ is the battery's maximum capacity in ampere-hour (Ah) and $I$ is the current in ampere ($I<0$ for discharging, and $I>0$ for charging). 
The second component of this model, $R_0$, represents the battery's internal resistance, which grasps the voltage drop in discharging and jump in charging. As the third component, a set of serially connected RC pairs    can characterize the transient behavior in the battery's voltage response. 
This work considers only one RC pair, which is often  sufficient practically and brings simplicity of analysis and computation. It is noteworthy that this will   cause no loss of generality as the ensuing results can be  extended to the case of multiple RC pairs.

By the Kirchhoff’s circuit laws, the voltage dynamics based on the one-RC Thevenin's model  
can be expressed as
\begin{subequations}
\begin{align}[left = \empheqlbrace\,]\label{Circuit-Model-VRC}
\dot V_{\mathrm{RC}} &  = -\frac{1}{RC}V_{\rm{RC}}-\frac{1}{C}I, \\ \label{Circuit-Model-Terminal-Voltage}
V & = V_{\mathrm{OC}}-V_{\rm{RC}}+R_0I,
\end{align}
\end{subequations}
where $V_{\rm RC}$ is the voltage across the RC pair,  $V_{\mathrm{OC}}$ the OCV, and $V$ the terminal voltage. 
As aforementioned, $V_{\rm OC}$ is dependent on SoC. Following~\citep{szumanowski2008battery,weng2014unified}, it can be characterized as a fifth-order polynomial  with respect to SoC:
\begin{align}\label{Equation-Voc}
V_{\mathrm{OC}}({\rm SoC})  = & \sum_{i=0}^5 \alpha_i \mathrm{SoC}^i, 
\end{align}
where $\alpha_i$ for $i = 0,1,\ldots,5$ are coefficients,  and $V_{\mathrm{OC}}$ is lower and upper bounded, respectively, by $\underline{V}_{\rm OC}=V_{\mathrm{OC}}({\rm SoC=0})$ and $\overline{V}_{\rm OC}=V_{\mathrm{OC}}({\rm SoC=1})$. In addition,  the literature often suggests  the internal resistance $R_0$  as a function of SoC,  which is in general constant but increases exponentially when SoC nears 0~\citep{chen2006accurate,lu2013review}. Hence,  $R_0$ is parameterized as
\begin{align}\label{Equation-R0}
R_0({\rm SoC}) =  \beta_0+\beta_1 e^{-\beta_2   \mathrm{SoC}},
\end{align}
where the coefficients $\beta_i>0$ for $i=0,1,2$.

From above, the Thevenin's model with one RC pair is represented by~\eqref{SoC-Counting}-\eqref{Equation-R0}. Now, suppose that  the  battery is fully charged,  left idling for a long period  and then discharged by a constant current to cut-off voltage. Here, the focus on a constant-current discharging scenario for the sake of formulating a mathematically tractable identification problem. As another advantage, such a discharging protocol can be easily implemented in practice. 
Solving~\eqref{Circuit-Model-VRC} with $V_{\rm RC}(t=0) = 0$, one can obtain the evolution of $V_{\rm RC}$  under constant current  through time as follows:
\begin{align}\label{VRC-behavior}
V_{\rm RC}(t) &=- IR\left(1-e^{-\frac{t}{RC}}\right).
\end{align}
As a result, the terminal voltage $V$ through time is 
\begin{align} \nonumber
V(t) = &\alpha_0+\alpha_1\mathrm{SoC}(t)+\alpha_2\mathrm{SoC}^2(t)+\alpha_3\mathrm{SoC}^3(t) +\alpha_4\mathrm{SoC}^4(t)+\alpha_5\mathrm{SoC}^5(t)\\ \label{Terminal-Voltage}
 &  +I\beta_0+I\beta_1 e^{-\beta_2 \mathrm{SoC}(t)}+IR\left(1-e^{-\frac{t}{RC}}\right) ,
\end{align}
where ${\rm SoC}(t) = 1+It/(3600Q_{\rm c})$ as indicated by~\eqref{SoC-Counting}, with ${\rm SoC}(0)=1$. 

 It is worth noting that $\underline{V}_{\rm OC}$ and $\overline{V}_{\rm OC}$ can be preset and used in the discharging experiments. It is also assumed that the total capacity $Q_c$ is known or can be experimentally determined for given $\underline{V}_{\rm OC}$ and $\overline{V}_{\rm OC}$.  
  Because  $\alpha_0=\underline{V}_{\rm OC}$ and $\sum^5_{i=0}\alpha_i=\overline{V}_{\rm OC}$,  only  $\alpha_i$ for $i=1,2,\ldots,4$ need to be determined. It is seen that~\eqref{Terminal-Voltage} defines an explicit relationship between  $V$,  $I$ and the unknown parameters, i.e.,  $\alpha_i$ for $i =1,\ldots,4$, $\beta_i$ for $i = 0,1,2$, $R$ and $C$. The next problem to tackle is two-fold: 1) deciding if the parameters are identifiable, and 2) designing approaches to achieve effective parameter identification, which will be the focus of the next section. Note that the parameters are assumed to be constant throughout this paper in order to consider  a representative and tractable problem. 
 

\section{Parameter Identification}\label{Sec:Parameter-Identification}


Based on  Section~\ref{Sec:Problem-Formulation}, this section aims to identify the unknown model parameters. It starts with investigating the parameter identifiability analysis, which shows that the parameters are locally identifiable and offers a method of determining  the theoretical identification accuracy. Then, it synthesizes two   identification methods, which are based on the notion of nonlinear least squares (NLS) but use constrained optimization and regularization, respectively,    to address the non-convexity issue that arises  in   parameter estimation.


\subsection{Identifiability Analysis}\label{Sec:Identifiability-Analysis}

Consider~\eqref{Terminal-Voltage}, for which the unknown parameter vector is 
\begin{align*}
{\bm \theta}=\left[\begin{matrix}\alpha_1 &\alpha_2&\alpha_3&\alpha_4&\beta_0&\beta_1&\beta_2&R&(RC)^{-1} \end{matrix}\right]^{\T}.
\end{align*}
Note that ${\theta}_i$ for $i = 1,2,\ldots,9$ and its corresponding parameter will be used interchangeably in sequel. 
The terminal voltage $V$ is measured at a sequence of sampling instants $\{t_k\}$ for $k=1,2,\ldots,N$ with $t_1=0$. By~\eqref{Terminal-Voltage}, $V(t_k)$  can be  compactly written as
\begin{align}\label{Terminal-Voltage-Compact-Equation}
V(t_k) = \phi(\bm \theta;t_k),
\end{align}
where 
\begin{align*} \nonumber
\phi(\bm \theta;t_k) =& \underline{V}_{\rm OC}+{\theta}_1\mathrm{SoC}(t_k)+{\theta}_2\mathrm{SoC}^2(t_k)+{\theta}_3\mathrm{SoC}^3(t_k)+{\theta}_4 \mathrm{SoC}^4(t_k) \\ \nonumber
 & +\left(\overline{V}_{\rm OC}-\underline{V}_{\rm OC}-\sum^{4}_{i=1}\theta_i\right) \mathrm{SoC}^5(t_k)+{\theta}_5I+{ \theta}_6 I e^{-{ \theta}_{7} \mathrm{SoC}(t_k)}
 +{ \theta}_8 I\left(1-e^{-{ \theta}_9 t_k}\right)  .
\end{align*}
Given~\eqref{Terminal-Voltage-Compact-Equation}, a fundamental question  is whether $\bm \theta$ can be uniquely identified from the data sequence $\{V(t_k)\}$, which boils down to the identifiability issue. To proceed, a definition of the local parameter identifiability is introduced for~\eqref{Terminal-Voltage-Compact-Equation} as follows~\citep{fang2014state,VanDoren:IFAC:SysID:2009}.
\begin{definition}\label{Definition:Sensitivity}
\setstretch{1.5}
A model structure $\phi(\bm \theta;t_k)$ is said to be locally identifiable at some point $\bm \theta^0$ in the parameter space for a given data sequence $\{V(t_1),V(t_2),$
$\ldots,V(t_N )\}$, if for any $\bm \theta^1$ and $\bm \theta^2$ within the neighborhood of $\bm \theta^0$, $\phi(\bm \theta^1;t_k)=\phi(\bm \theta^2;t_k)$ holds if and only if $\bm \theta^1=\bm \theta^2$.
\end{definition}
 Definition~\ref{Definition:Sensitivity} indicates that there should exist no two different parameter sets in the neighborhood of a point that can produce the same data sequence if the parameters are locally identifiable. 

Before moving on to discuss the identifiability testing condition, let us consider the parameter identification problem for~\eqref{Terminal-Voltage-Compact-Equation}  in a prediction-error framework, seeking to find out the  parameter estimates to minimize the difference between the measurements and the  model-based predictions. 
To this end, an NLS cost function can be formulated as
\begin{align}\label{Prediction-error-minimization}
J(\bm \theta) = {1 \over 2} \left[\bm y-{\bm \phi}(\bm\theta)\right]^{\T} {\bm Q}^{-1}\left[\bm y-{\bm \phi}(\bm\theta)\right],
\end{align}
where $\bm Q$ is a symmetric positive definite matrix interpretable as the covariance matrix of an additive measurement noise, and
\begin{align*}
{\bm y}  &= \begin{bmatrix} V(t_1) & V(t_2) & \cdots & V(t_N) \end{bmatrix}^{\T},\\
{\bm \phi}({\bm{\theta}})& = \begin{bmatrix} \phi(\bm \theta;t_1) & \phi(\bm \theta;t_2) & \cdots & \phi(\bm \theta;t_N) \end{bmatrix}^{\T}.
\end{align*}
The parameter estimation then lies in finding out $\hat{\bm \theta}$ to minimize $J(\bm \theta)$, i.e.,
\begin{align}\label{Basic-Identification}
\hat{\bm \theta} = \arg\min_{\bm \theta} J(\bm \theta).
\end{align}
It is pointed out in~\citep{VanDoren:IFAC:SysID:2009} that evaluating the local identifiability of $\bm \theta$ around $\hat{\bm \theta}$ (see Definition~\ref{Definition:Sensitivity}) can be achieved by testing whether the minimization problem in~\eqref{Prediction-error-minimization} has a unique solution in the local parameter space. Further, a sufficient condition to guarantee a unique $\hat{\bm \theta}$ (or in other words, the local identifiability of $\bm \theta$) is that  the Hessian matrix $\partial^2 J({\bm \theta}) /\partial {\bm \theta}^2 > \bf 0$ at $\hat{\bm \theta}$. To satisfy this condition, $\partial {\bm \phi} ({\bm \theta}) /\partial {\bm \theta}$ at $\hat{\bm \theta}$ must be of full rank. This is because the local first-order approximation of $\partial^2 J({\bm \theta}) /\partial {\bm \theta}^2$ is 
\begin{align*}
\frac{\partial^2 J({\bm \theta})}{\partial {\bm \theta}^2} \approx \left(\frac{\partial {\bm \phi}({\bm \theta}) }{\partial {\bm \theta}}\right)^\T {\bm Q}^{-1} \frac{\partial {\bm \phi}({\bm \theta}) }{\partial {\bm \theta}} .
\end{align*}
Here, define ${\bm S}({\bm \theta}) = \partial {\bm \phi} ({\bm \theta}) /\partial {\bm \theta}$. It is referred to as the sensitivity matrix and measures how  the parameter perturbations can change the terminal voltage. The terminal voltage can be said to be sensitive to a parameter if it sees a relatively large change when the parameter varies slightly. A parameter with higher sensitivity generally allows for more accurate estimation. Meanwhile,  ${\bm S}({\bm \theta})$ should be of full rank as discussed above, so that when the parameters change, their effects on the terminal voltage are differentiable.
Specifically, ${\bm S}({\bm \theta})$ is given by
\begin{align}\label{Sensitivity_matrix}
{\bm S}({\bm \theta}) = \left[ \begin{array}{cccc}
\vdots & \vdots  & \vdots & \vdots\\
 \dfrac{\partial { \phi}({\bm \theta};t_k)}{\partial  { \theta}_1}&\dfrac{\partial { \phi}({\bm \theta};t_k)}{\partial  { \theta}_2}
 & \cdots & \dfrac{\partial { \phi}({\bm \theta};t_k)}{\partial  { \theta}_{9}}\\
\vdots & \vdots  & \vdots & \vdots\\
\end{array} \right]_{N\times 9}, 
\end{align}
where 
\begin{subequations}\label{Sensitivity_for_each_parameter}
\begin{align}\label{Sensitivity_1_to_4}
\frac{\partial { \phi}({ \bm \theta};t_k)}{\partial  { \theta}_i}  = & \mathrm{SoC}^i(t_k)-\mathrm{SoC}^5(t_k),~{\textrm{for}}~i = 1,\cdots,4,\\ \label{Sensitivity_5}
\frac{\partial { \phi}({ \bm \theta};t_k)}{\partial  { \theta}_{5}}   = &I,\\ \label{Sensitivity_6}
\frac{\partial { \phi}({ \bm \theta};t_k)}{\partial  { \theta}_{6}}   = & I e^{-{ \theta}_{7}\mathrm{SoC}(t_k)   },\\ \label{Sensitivity_7}
\frac{\partial { \phi}({\bm  \theta};t_k)}{\partial  { \theta}_{7}}   = & -I\mathrm{SoC}(t_k)  { \theta}_{6}  e^{-{ \theta}_{7}\mathrm{SoC}(t_k)}, \\   \label{Sensitivity_8}
\frac{\partial { \phi}({\bm  \theta};t_k)}{\partial  { \theta}_8}  = & I\left(1-e^{-{ \theta}_{9}{t_k} }\right),\\ \label{Sensitivity_9}
\frac{\partial { \phi}({ \bm \theta};t_k)}{\partial  { \theta}_9}  = & I {\theta}_{8}t_ke^{-{ \theta}_{9}{t_k}}, 
\end{align}
\end{subequations}
for $k = 1,2,\cdots,N$. From above one can observe that the columns of ${\bm S}({\bm \theta})$ are  linearly independent, making ${\bm S}({\bm \theta})$ a full-rank matrix. This conclusion can by further verified by computationally evaluating  ${\bm S}({\bm \theta})$ around the nominal parameters of a battery model, with further discussion offered in Section~\ref{Sec:Numerical-Simulation}. Hence, one can claim that the parameter vector $\bm \theta$ is   locally identifiable.  

Further, suppose that $\hat{\bm\theta}$ is successfully estimated and minimizes $J(\bm\theta)$. The covariance of $\hat{\bm\theta}$ in the Gaussian case then is given by
\begin{align}\label{Covariance-Parameter}
{\rm Cov}({\hat{\bm \theta}}) =  {\rm E}\left(\left[ \left. \frac{\partial^2 J ({\bm \theta})}{\partial {\bm \theta}^2} \right|_{{\bm \theta}} \right]^{-1}\right) \approx  \left[\bm S^{\T}({\bm \theta})\bm Q^{-1}  \bm S({\bm \theta})\right]^{-1},
\end{align}
from which the variance of the estimate $\hat{\theta}_{i}$ for $i=1,2,\cdots,9$, is the $i$-th diagonal element of ${\rm Cov}({\hat {\bm \theta}})$, i.e., $\left[{\rm Cov({\hat{\bm \theta}})} \right]_{ii}$. Here, it should be noted that $\left[{\rm Cov({\hat{\bm \theta}})} \right]_{ii}$ measures the estimation error if the estimation is unbiased and one can use~\eqref{Covariance-Parameter} to calculate the theoretically possible parameter estimation accuracy.

\subsection{Identification Methods}\label{Sec:Identification-Methods}

The foregoing discussion poses a basic parameter estimation problem as shown in~\eqref{Prediction-error-minimization}-\eqref{Basic-Identification}. Although the parameters are   locally identifiable, solving the problem is still a challenge due to   two  difficulties. The first one stems from the nonlinearity of the model, as is seen from the voltage equation~\eqref{Terminal-Voltage}. As the second and more challenging difficulty, the minimization problem in~\eqref{Basic-Identification} is non-convex. Although a numerical optimization procedure, e.g., the Gauss-Newton method, can be deployed to tackle the nonlinearity issue, the non-convexity may still lead the parameter search to local minimum points that are unphysical or unreasonable. To ensure correct parameter estimation, this work introduces constrained optimization and regularization to improve the problem formulation by using prior knowledge about parameters. 

\subsubsection{Constrained Optimization}\label{Sec:Constrained-Optimization}
To prevent parameter search approaching  physically meaningless local minima, one can constrain the search within a  parameter space believably correct. Specifically, one can roughly determine the lower and upper bounds of some parameters, use them to set up a limited search space and run numerical optimization within this space. In practice, it is not difficult to determine the bounds of some parameters  for the Thevenin's model, because some coarse-grained knowledge of a battery, e.g., internal impedance, can be obtained from both experience and some simple observation or analysis of the measurement data. A further discussion of extracting prior knowledge of parameters is provided in Section~\ref{Sec:Experimental-Validation}. With this idea, the identification problem in~\eqref{Basic-Identification} is modified as a constrained optimization problem:
\begin{subequations}\label{CNLS-cost-problem}
\begin{align}\label{objective-function}
\hat{\bm \theta} =  \min_{\bm \theta}  \  J(\bm \theta), \\ \label{constraint}
 \textrm{s.t.} \  \underline{\bm\theta} \leq \bm \theta \leq \overline{\bm\theta },
\end{align}
\end{subequations}
where $\underline{\bm\theta}$ and $ \overline{\bm\theta }$ are the  lower and upper bounds for $\bm\theta$, respectively. It is noted that~\eqref{CNLS-cost-problem} represents a constrained NLS problem, which can be addressed by the  trust region or line search algorithms. Here, a trust region method is considered, and a brief overview about it is taken from~\citep{coleman1996interior} and offered below for completeness. 

Let us begin with the unconstrained case and consider~\eqref{CNLS-cost-problem}  without the constraint~\eqref{constraint}. Suppose there is a current guess of the solution $\hat{\bm \theta}_k$, where $k$ is the iteration step number.  For the trust region method, it first constructs a  function $\psi_k$,   referred to as ``model function'', to approximate the actual objective function $J$ around the current guess $\hat{\bm \theta}_k$. The model function $\psi_k$ is based on the Taylor-series expansion of $J$ around $\hat{\bm \theta}_k$ and defined as follows:
\begin{align}\label{trust-region-approximate-model}
\psi_k(\bm s_k)=  J(\hat{\bm \theta}_k)+ \bm g_k^{\T}\bm s_k+\frac{1}{2}\bm s_k^{\T}\bm B_k \bm s_k ,
\end{align}
where $\bm s_k$ is a step to be determined, $\bm g_k=\nabla J ( \hat{\bm \theta}_k ) $, and $\bm B_k=\nabla^2 J ( \hat{\bm \theta}_k ) $. Then, minimizing the objective function $J$ reduces to finding a minimizer of the model function $\psi_k(\bm s_k)$. Since $\psi_k(\bm s_k)$ can be a poor approximation of $J$ when the step $\bm s_k$ is too large, the trust region method bounds the search for a minimizer of $\psi_k(\bm s_k)$ within some region around $\hat{\bm \theta}_k$. Hence,  one needs to solve the following subproblem in order to find out the best candidate step $\bm s_k$:
\begin{align}\label{subproblem-function}
\min_{ {\bm s}_k } \  \psi_k(\bm s_k)~~~
\textrm {s.t.}\ \lVert \bm D_k \bm s_k  \lVert \leq {\Delta}_k,
\end{align}
where $\bm D_k$ is a scaling matrix, $\Vert \cdot \Vert$ is the Euclidean norm, and $\Delta_k$ is the so-called trust region radius. It is then interesting to see whether $\bm s_k$ can bring a large enough drop in $J(\bm\theta)$ if added to $\hat{\bm \theta}_k$. A worthy metric for this is
\begin{align}\label{gain}
\rho_k = \frac{   J(\hat{\bm \theta}_k) -J (\hat{\bm \theta}_k+ \bm s_k)}{\psi_k( 0)-\psi_k( \bm s_k) },
\end{align}
where $J(\hat{\bm \theta}_k) -  J(\hat{\bm \theta}_k+ \bm s_k)$ is the actual reduction in $J(\bm\theta)$, and $\psi_k( 0)-\psi_k( \bm s_k)$ the anticipated reduction. If $\rho_k$ is greater than a threshold, it is safe to increase the trust-region radius by letting $\hat{\bm \theta}_{k+1} = \hat{\bm \theta}_k+\bm s_k$ and then continuing the search. Otherwise, the trust region should be reduced, and rerun~\eqref{subproblem-function} at the same iterate point $\hat{\bm \theta}_k$.

In the case of~\eqref{constraint} applied as a constraint, there are different ways to cope with this. An approach offered in~\citep{coleman1996interior} defines a new subproblem, in which $\psi_k(\bm s_k)$ is replaced by
\begin{align*}\label{new-subproblem}
\psi'_k(\bm s_k) = J(\hat{\bm \theta}_k)+ \bm g_k^{\T}\bm s_k+\frac{1}{2}\bm s_k^{\T}(\bm B_k +\bm C_k )\bm s_k,
\end{align*}
where $\bm C_k$ is a matrix that results from the constraint.  Similar to~\eqref{trust-region-approximate-model}, this subproblem is then iteratively implemented to search for the optimal parameter estimates. Summarizing the above, one can obtain the constrained  NLS method  for the Thevenin's model identification,  which is named as C-NLS. 
\subsubsection{Regularization}\label{Sec:Regularization}

Tikhonov regularization offers another way to help overcome the non-convexity issue. It involves a pre-estimation of the parameters  and  uses the prior guess to regularize the original cost function. Here, the pre-estimation is expected to be close to the truth so that its presence will then drive the optimization to run in the vicinity of the true parameter values. The regularization-based parameter estimation problem is expressed as follows:
\begin{align}\label{RNLS-cost-problem}
\min_{\bm \theta } \ \bar J (\bm \theta)=  J({\bm \theta})+ {1 \over 2}\left( \bm \theta-\bm \theta_0 \right)^{\T}{\bm P_0^{-1}}\left( \bm \theta-\bm \theta_0 \right),
\end{align}
where $\bm \theta_0 $ represents the prior guess of $\bm \theta$, and $\bm P_0 $ is a diagonal positive matrix acting as a quantification of the confidence in the quality of $\bm \theta_0 $. This formulation in~\eqref{RNLS-cost-problem} can be interpreted from a Bayesian perspective, where $\bm\theta$ is assumed to be a Gaussian random vector following $\mathcal{N}({\bm \theta}_0, {\bm P}_0)$. More information about this Bayesian-perspective interpretation can be found in Appendix. Here, the selection of $\bm \theta_0 $ and $\bm P_0 $ depends on some prior knowledge of a battery under consideration, and further discussion is given in Section~\ref{Sec:Experimental-Validation}. To address the problem in~\eqref{RNLS-cost-problem}, one can also consider the trust region method introduced in Section~\ref{Sec:Constrained-Optimization}. Thus far  the regularization-based NLS method is obtained, which is named as R-NLS.

\subsubsection{Comparison of C-NLS and R-NLS}\label{Sec:Comparison-C-R-NLS}
Both  the C-NLS and R-NLS methods are developed  to accomplish the  parameter estimation for the Thevenin's model. Sharing the  objective of overcoming the non-convexity issue, they  both  require some   advance knowledge of the parameters (i.e., the parameter bounds $\underline{\bm\theta}$ and $\overline{\bm\theta}$ for C-NLS and prior guess $\bm\theta_0$ and $\bm{P}_0$ for R-NLS). Our experience suggests that the advance knowledge does not have to be accurate---even a coarse inference   can lead to satisfactory estimation accuracy. It also turns out that one can gather it in some convenient ways, with a further discussion to be offered in Section~\ref{Sec:Experimental-Validation} along with the elaboration of    experiments. 

Meanwhile, there are several interesting differences between the two approaches. The first one lies in the computational efficiency. It is noteworthy that the R-NLS is usually computationally faster as it is based on unconstrained optimization. A second difference is concerned with the parameter estimation accuracy. Based on~\cite{hoerl1970ridge},  the theoretical estimation error for both approaches can be quantified as $\tr(\bm\Sigma)$, where $\tr$ denotes trace and 
\begin{align}\label{Sigma-Error}
{\bm\Sigma}=
\mathrm{E}\left[(\hat{\bm\theta}-\bm\theta)(\hat{\bm\theta}-\bm\theta)^{\T}\right]
={\rm Cov}(\hat{\bm\theta})+\left(\mathrm{E}[\hat{\bm\theta}]-{\bm\theta}\right)
\left(\mathrm{E}[\hat{\bm\theta}]-{\bm\theta}\right)^{\T}.
\end{align}
Here, one can approximately view $[ \Sigma]_{ii}$ as a squared error between $\hat\theta_i$ and $\theta_i$, and the second term of the rightmost-hand side is due to the bias of  $\mathrm{E}[\hat{\bm\theta}]$ from ${\bm\theta}$. 
For the C-NLS method,   the bias term in~\eqref{Sigma-Error} will reduce to  zero if the bounds are ideally selected. In view of~\eqref{Covariance-Parameter}, its theoretical estimation accuracy then is given by 
\begin{align}\label{CSigma-Accuracy}
\tr(\bm\Sigma)=\tr\left({\left[\bm S^{\T}({\bm \theta})\bm Q^{-1}  \bm S({\bm \theta})\right]^{-1}}\right).
\end{align}
 By contrast, the R-NLS method produces biased estimates as it pushes the estimates toward $\bm\theta_0$. It follows from~\eqref{RNLS-cost-problem} that
\begin{align*}\nonumber
\mathrm{E}[\hat{\bm\theta}]=\bm\theta+\left(\bm{I}+\bm{P}_0\bm{S}^{\T}({\bm \theta})\bm{Q}^{-1}\bm{S}({\bm \theta})\right)^{-1}(\bm\theta-\bm\theta_0).
\end{align*}
This hence suggests that the theoretical estimation error of the R-NLS is
\begin{align}\label{RSigma-Accuracy}
\tr(\bm\Sigma)=\tr{\left(\left[\bm S^{\T}({\bm \theta})\bm Q^{-1}  \bm S({\bm \theta})+\bm{P}_0^{-1}\right]^{-1}\right)}
+(\bm\theta-\bm\theta_0)^{\T}
\left(\bm{I}+\bm{P}_0\bm{S}^{\T}({\bm \theta})\bm{Q}^{-1}\bm{S}({\bm \theta})\right)^{-\T}
\left(\bm{I}+\bm{P}_0\bm{S}^{\T}({\bm \theta})\bm{Q}^{-1}\bm{S}({\bm \theta})\right)^{-1}
(\bm\theta-\bm\theta_0).
\end{align}
Comparing~\eqref{CSigma-Accuracy} and~\eqref{RSigma-Accuracy}, it is easy to notice the theoretical estimation accuracy attained by the two methods can be different, with the former being fixed and the latter dependent on $\bm\theta_0$ and $\bm{P}_0$. With distinct choices of $\bm\theta_0$ and $\bm{P}_0$, the R-NLS can be either more accurate than the C-NLS (e.g., $\bm\theta_0=\bm\theta$ and $\bm{P}_0=\mathbf{0}$ as an extreme  case) or less accurate (e.g., $\bm\theta_0$ is far away from $\bm\theta$ and $\bm{P}_0$ is close to $\bm0$). This indicates that one can leverage~\eqref{RSigma-Accuracy} as a guidance for the selection of $\bm\theta_0$ and $\bm{P}_0$, and a detailed discussion of this will be given in Section~\ref{Sec:Numerical-Simulation}. 

\section{Monte Carlo Simulation}\label{Sec:Numerical-Simulation}
This section presents Monte Carlo simulation to verify the effectiveness of the C-NLS and R-NLS methods in extracting the Thevenin's model parameters. The simulation will further compare them in terms of estimation accuracy and computation time in accordance with the analysis in Section~\ref{Sec:Comparison-C-R-NLS}.

\subsection{Training Data}
The training data used in the Monte Carlo simulation are generated using the Thevenin's model with the following parameters: the nominal capacity $Q_{\rm c}=2.17~{\rm Ah}$, $R=0.0313~\Omega$,  $C=1,858~\rm F$, and 
\begin{align}  \nonumber
R_0({\rm SoC}) = &  \ 0.0313+0.0678 \cdot e^{-13.2 \cdot \mathrm{SoC}}, \\ \nonumber
V_{\mathrm{OC}}({\rm SoC})  = &   \ 3.3+2.61 \cdot \mathrm{SoC}-9.36\cdot\mathrm{SoC}^2  +19.7\cdot \mathrm{SoC}^3 
 -19.0\cdot \mathrm{SoC}^4+6.9 \cdot \mathrm{SoC}^5,
\end{align}
with $\underline{V}_{\rm OC}=3.3~\rm V$ and $\overline{V}_{\rm OC}=4.15~\rm V$. Here, the battery parameters are drawn  from the identification results on a Samsung INR18650-25R Li-ion battery used in Section~\ref{Sec:Experimental-Validation}.
  
\begin{figure}[t]
\centering
{\includegraphics[trim = {5mm 6mm 10mm 14mm}, clip, width=0.42\textwidth]{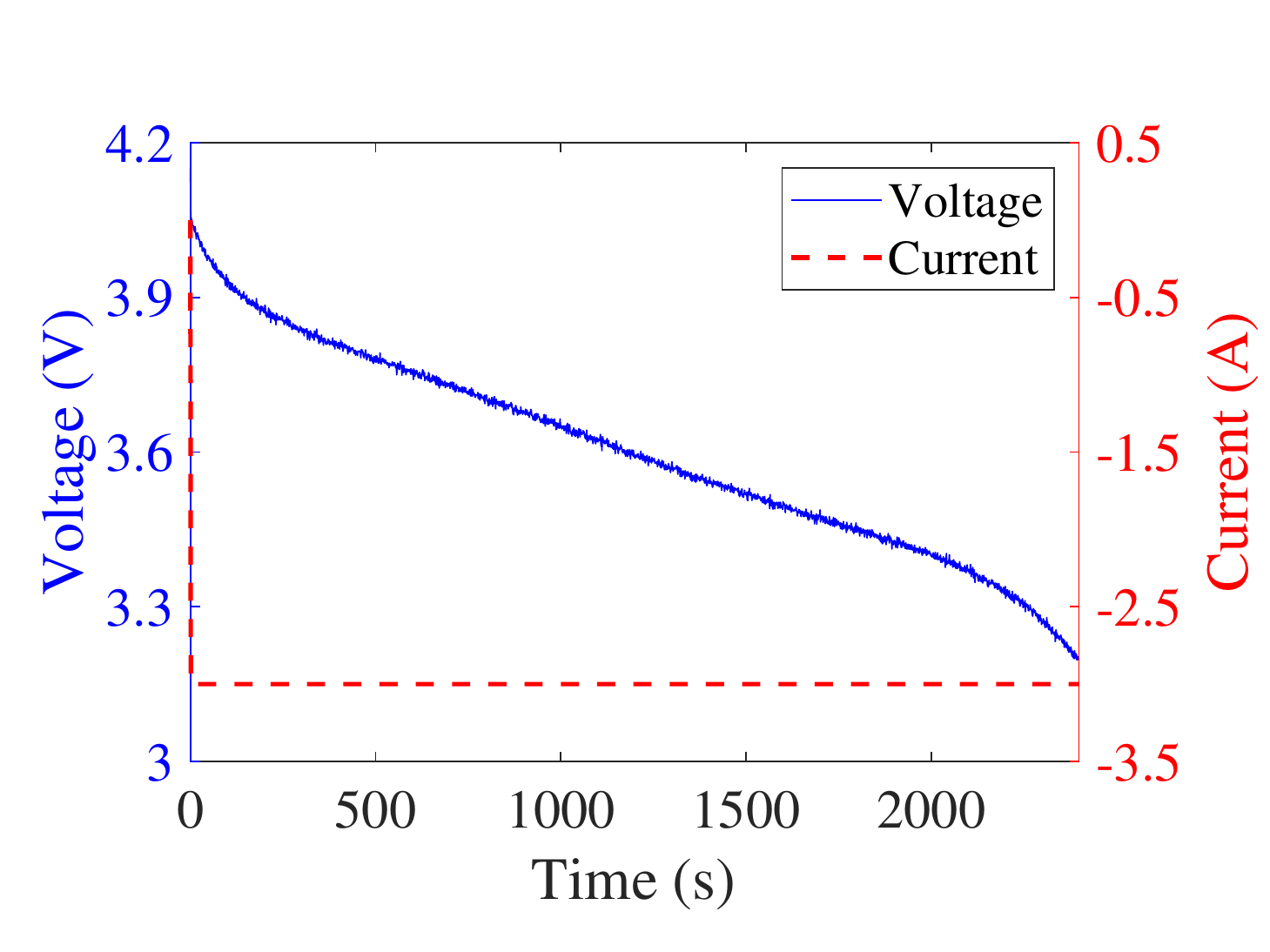}}
\caption{A model-based computer-generated terminal voltage profile for one Monte Carlo simulation.} 
\label{Fig:Cons-IV-Simulation}
\vspace{0mm}
\end{figure}

Suppose the battery is discharged from full charge under a constant current $I$ of $-3$ A by 2,400 s until when the terminal voltage reaches  the cut-off voltage of 3.2 V. By~\eqref{SoC-Counting}, the SoC will decrease from 1 following
\begin{align}\label{SoC-Counting-Simulation}
\mathrm{SoC}(t) & = 1-\frac{1}{2604}t.
\end{align}
According to~\eqref{Terminal-Voltage}, the terminal voltage through time is given by
\begin{align}  \label{Terminal-Voltage-Simulation} \nonumber
V(t)  = &~  3.3+{\theta}_{1}\cdot \mathrm{SoC}(t) +{ \theta}_{2}\cdot\mathrm{SoC}^2(t)  +{ \theta}_{3}\cdot\mathrm{SoC}^3(t) +{ \theta}_{4}\cdot\mathrm{SoC}^4(t)\\ 
& ~+\left( 4.15-3.3-\sum^4_{i=1}\theta_i\right)\cdot\mathrm{SoC}^5(t) 
-{\theta}_{5}\cdot3- {\theta}_{6}\cdot 3 e^{-{\theta}_{7}\cdot {\mathrm{SoC}}(t)}
-{\theta}_{8}\cdot 3\left(1-e^{-{\theta}_{9}\cdot t}\right),
\end{align}
where ${\theta}_i$ for $i=1,2,\ldots,9$ make up the parameter vector $\bm\theta$ with
\begin{align*}
{\bm \theta}=\left[\begin{matrix} 2.61  & -9.36  & 19.7 &  -19.0
&0.0313  &  0.0678&   13.2&  0.0313  &0.0172 \end{matrix} \right]^{\T} .
\end{align*}
Let the terminal voltage $V(t)$ be sampled every second and corrupted by an additive white Gaussian noise with variance of $2.5\times10^{-5}$ (i.e., $\bm Q=2.5\times10^{-5}\bm I$ in~\eqref{CNLS-cost-problem} and~\eqref{RNLS-cost-problem}). Running~\eqref{SoC-Counting-Simulation}-\eqref{Terminal-Voltage-Simulation}, one can obtain an $I$-$V$ dataset, with an example  plotted in Figure~\ref{Fig:Cons-IV-Simulation}. Further, repeat the same procedure for $M=500$ times to create 500 datasets. These datasets, accounting for the effect of random noise sampling, can help develop  statistically convincing conclusions about the performance of the proposed methods.

Before proceeding to parameter identification, one can first calculate the parameter sensitivity matrix ${\bm S}({\bm \theta})$ based on the datasets to computationally verify the full rankness of  ${\bm S}({\bm \theta})$. It can be easily and consistently found out that ${\bm S}({\bm \theta})$ is of full rank, showing that  $\bm \theta$ is locally identifiable as previously analyzed.

\subsection{Simulation Results}\label{Sec:MC-Simulation}


With the datasets generated and the parameters' local identifiability  verified, let us now apply the C-NLS and R-NLS methods to each dataset for parameter estimation. Meanwhile, the original NLS problem~\eqref{Prediction-error-minimization} is also solved to provide a benchmark for comparison. Table~\ref{Tab:Parameters-Setting} summarizes the initial guess for all the methods, $\underline{\bm\theta}$ and $\overline{\bm\theta}$ for the C-NLS, and $\bm\theta_0$ and $\bm P_0$ for the R-NLS. Note that the setting in Table~\ref{Tab:Parameters-Setting}  is coarse-grained relative to the truth.
 Here, consider the normalized root-mean-square error (NRMSE) to quantify the parameter estimation accuracy, which is defined as
\begin{align}\label{NRMSE}
 {\rm NRMSE}(\theta_i) =\sqrt{
 \frac{{1 \over M} \sum^M_{k=1} 
 \left( \hat{\theta}_i[k] -  {\theta}_{i} \right)^2}{\theta_i^2}
 },
\end{align}
for $ i = 1, 2, \ldots, 9$. Recalling~\eqref{CSigma-Accuracy}-\eqref{RSigma-Accuracy}, one can notice that ${\rm NRMSE}(\theta_i)$ is a sample-based computational estimation of  $\sqrt{[\Sigma]_{ii}/\theta_i^2}$.  They are expected to be close if $M$ is large enough.

\begin{table}[t]\centering
\begin{threeparttable}
\caption{The simulation setting.}
\begin{tabular}{ccccccccccc}
\toprule%
 &Par&$\alpha_1/1$&$\alpha_2/1$ &$\alpha_3/1$ & $\alpha_4/1$  & $\beta_0/\Omega$    & $\beta_1/\Omega$    & $\beta_2/1$ & $R/\Omega$   & $(RC)^{-1}/{\rm s}^{-1}$  \\
 \midrule
& $\rm{Initial~Guess}$&1  &  1 &   1 &   1  & 0.029  & 0.4 &   40&  0.2  &  1/40  \\
C-NLS&$\underline{\bm\theta}$ &- & - & - & -  &  0.01  & 0  &  0 &  0  &  1/200   \\
&$\overline{\bm\theta}$ &- & - & - & -  &  0.04  & 0.8 & 80   &  0.4  &  1   \\
R-NLS&$\bm \theta_0$ & 1 & 1 & 1 & 1  &  0.029  & 0.4 &  40 & 0.2 &  1/40   \\
& ${\rm{diag}}[\bm P_0]$ & $50^2$ &$50^2$&$50^2$ &$50^2$  &  $0.001^2$  & $0.1^2$ &  $10^2$ &  $0.06^2$  & $0.005^2$ \\
\bottomrule
\end{tabular}
\label{Tab:Parameters-Setting}
\end{threeparttable}
\end{table}

\begin{figure}[h]
\centering
\subfigure[NLS as a benchmark]
{\includegraphics[trim = {0mm 0mm 0mm 0mm}, clip, width=0.3\textwidth]{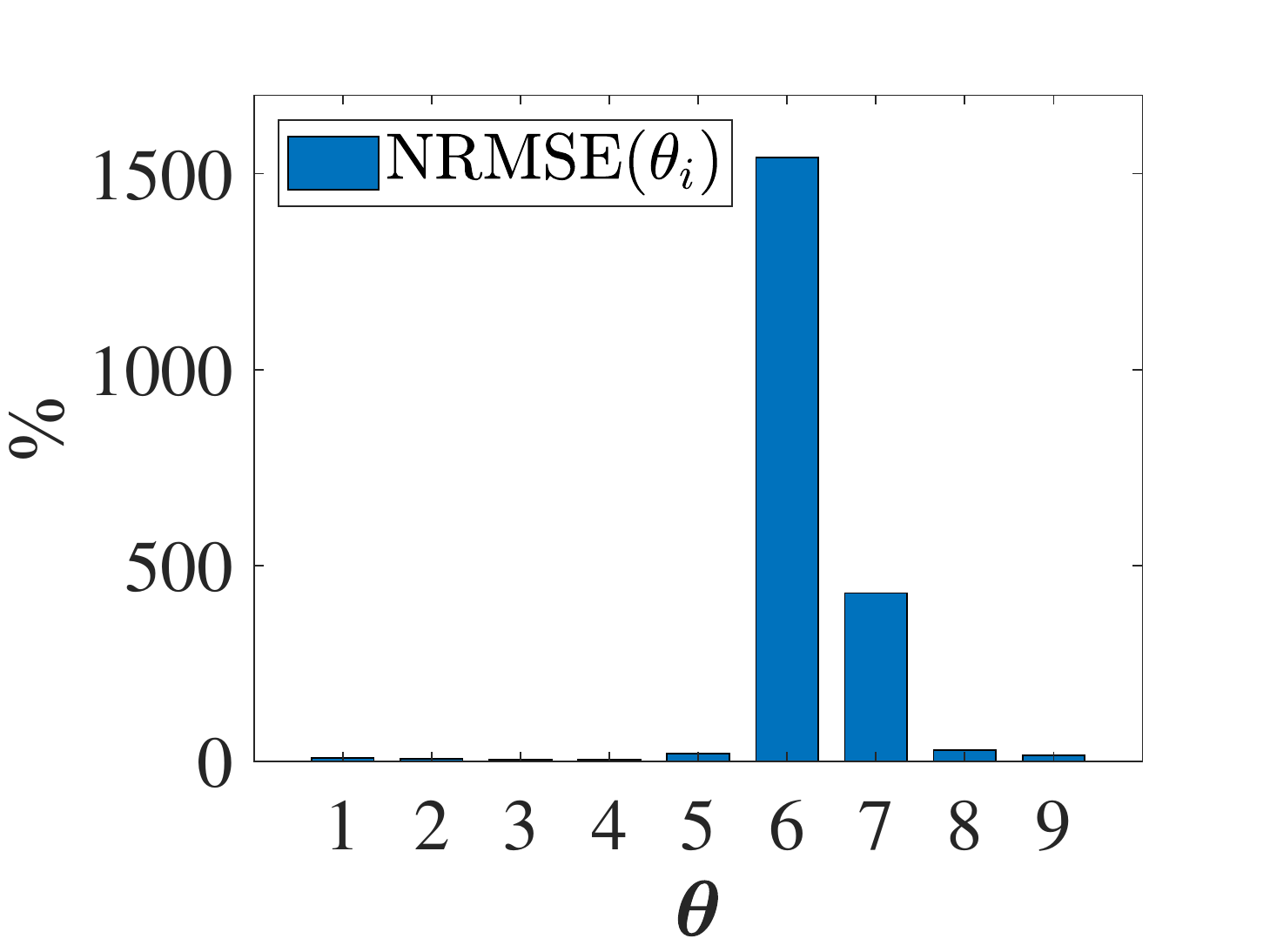}\label{Fig:Covariance-NLS}}
\subfigure[C-NLS]
{\includegraphics[trim = {0mm 0mm 0mm 0mm}, clip, width=0.3\textwidth]{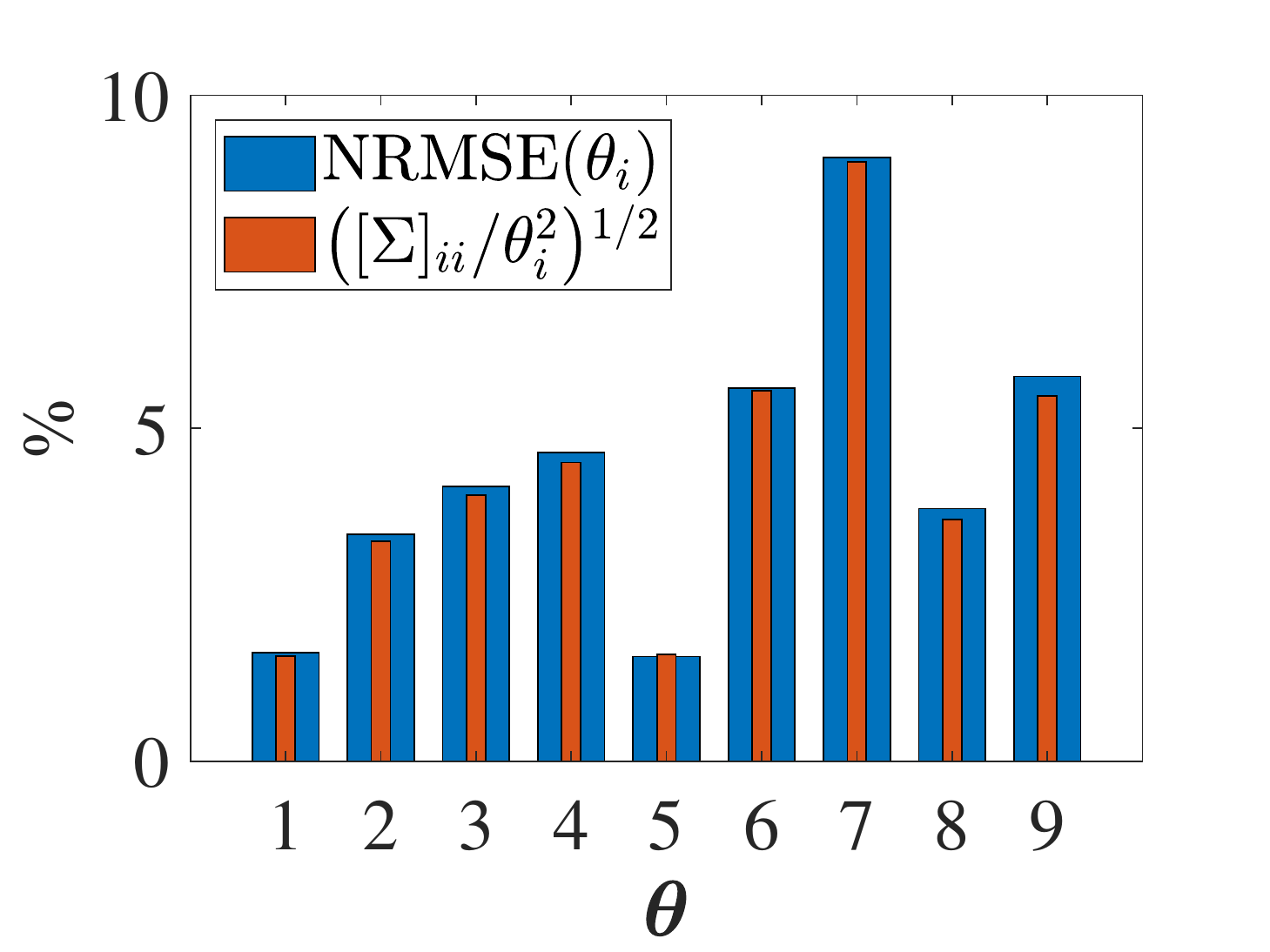}\label{Fig:Covariance-CNLS}}
\subfigure[R-NLS]
{\includegraphics[trim = {0mm 0mm 0mm 0mm}, clip, width=0.3\textwidth]{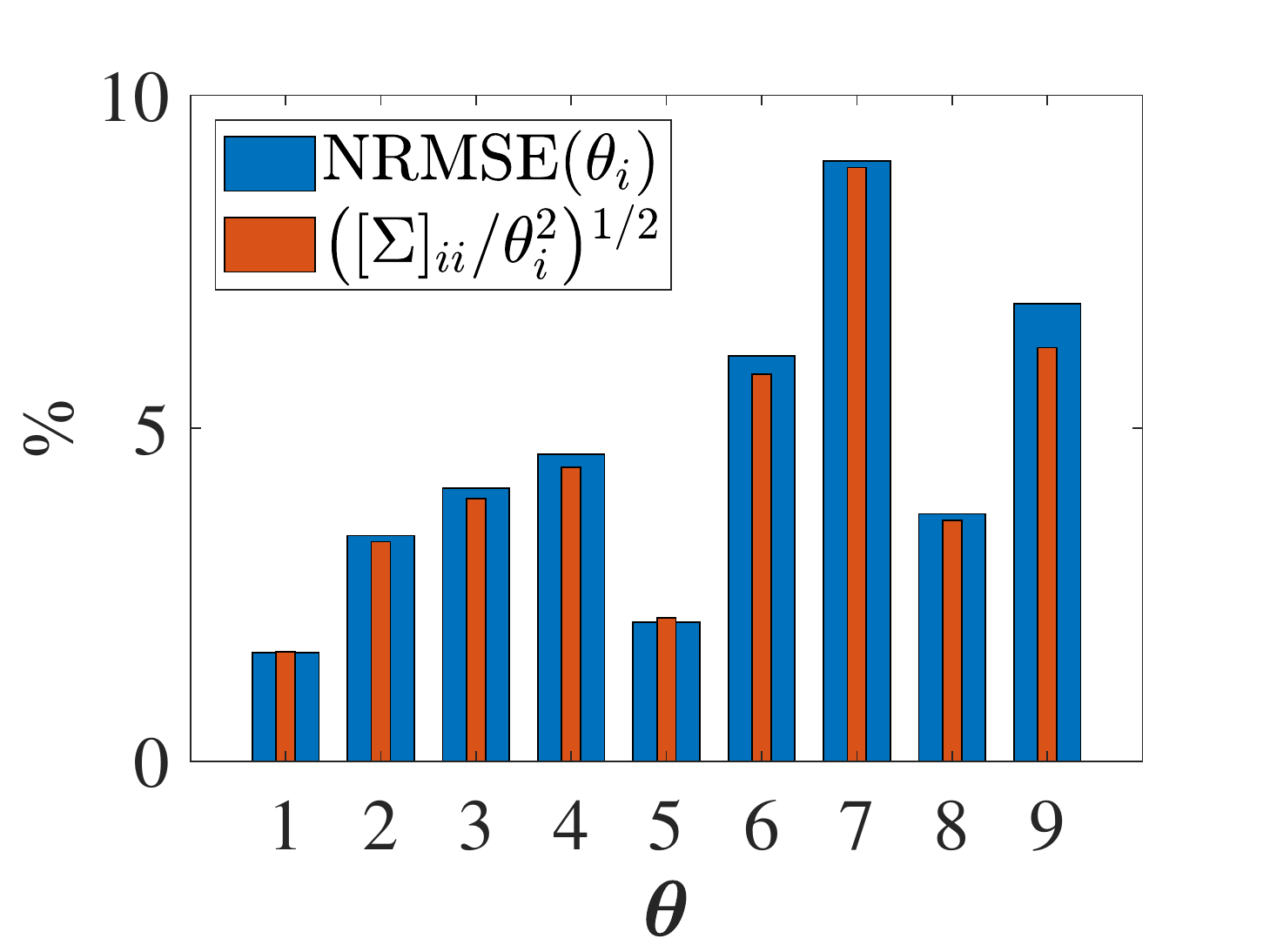}\label{Fig:Covariance-RNLS}}
\centering
\caption{Parameter estimation errors for (a) NLS as a benchmark, (b) C-NLS and (c) R-NLS.} 
\label{Fig:3NLS-Covariance}
\vspace{0mm}
\end{figure}

\begin{figure}[h]
\centering
\subfigure[NLS as a benchmark]
{\includegraphics[trim = {0mm 0mm 0mm 0mm}, clip, width=0.3\textwidth]{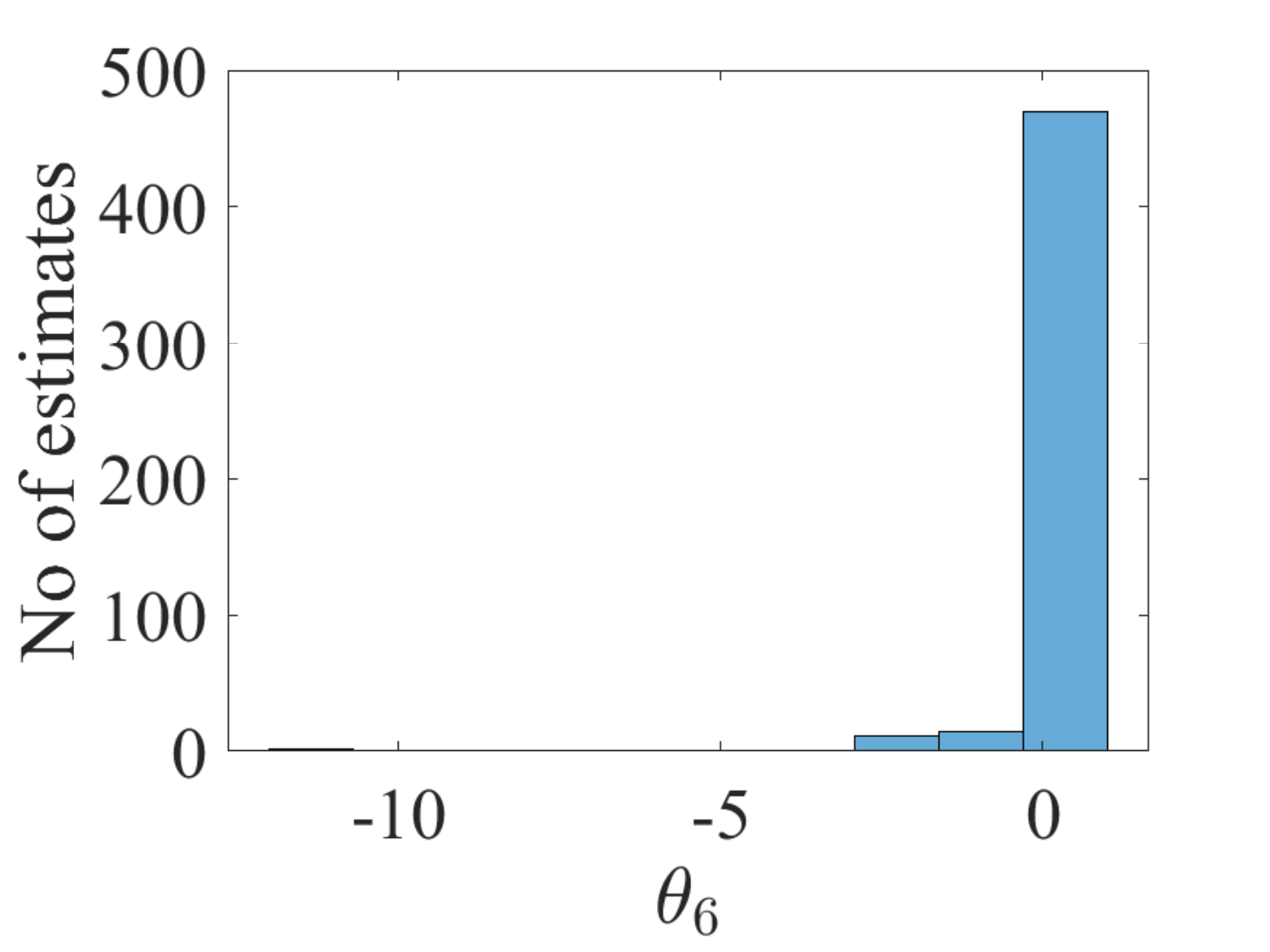}\label{Fig:Hist-NLS}}
\subfigure[C-NLS]
{\includegraphics[trim = {0mm 0mm 0mm 0mm}, clip, width=0.3\textwidth]{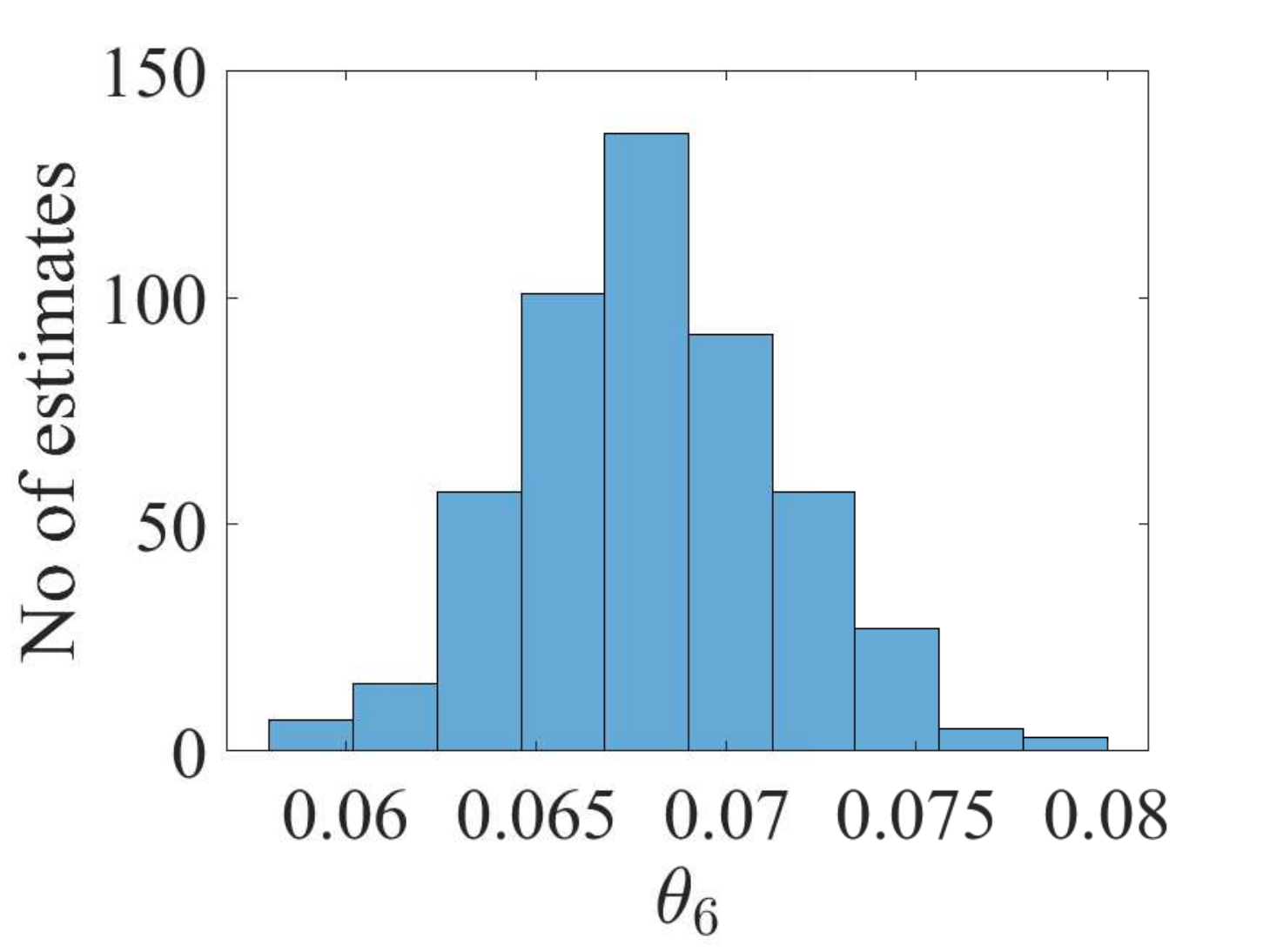}\label{Fig:Hist-CNLS}}
\subfigure[R-NLS]
{\includegraphics[trim = {0mm 0mm 0mm 0mm}, clip, width=0.3\textwidth]{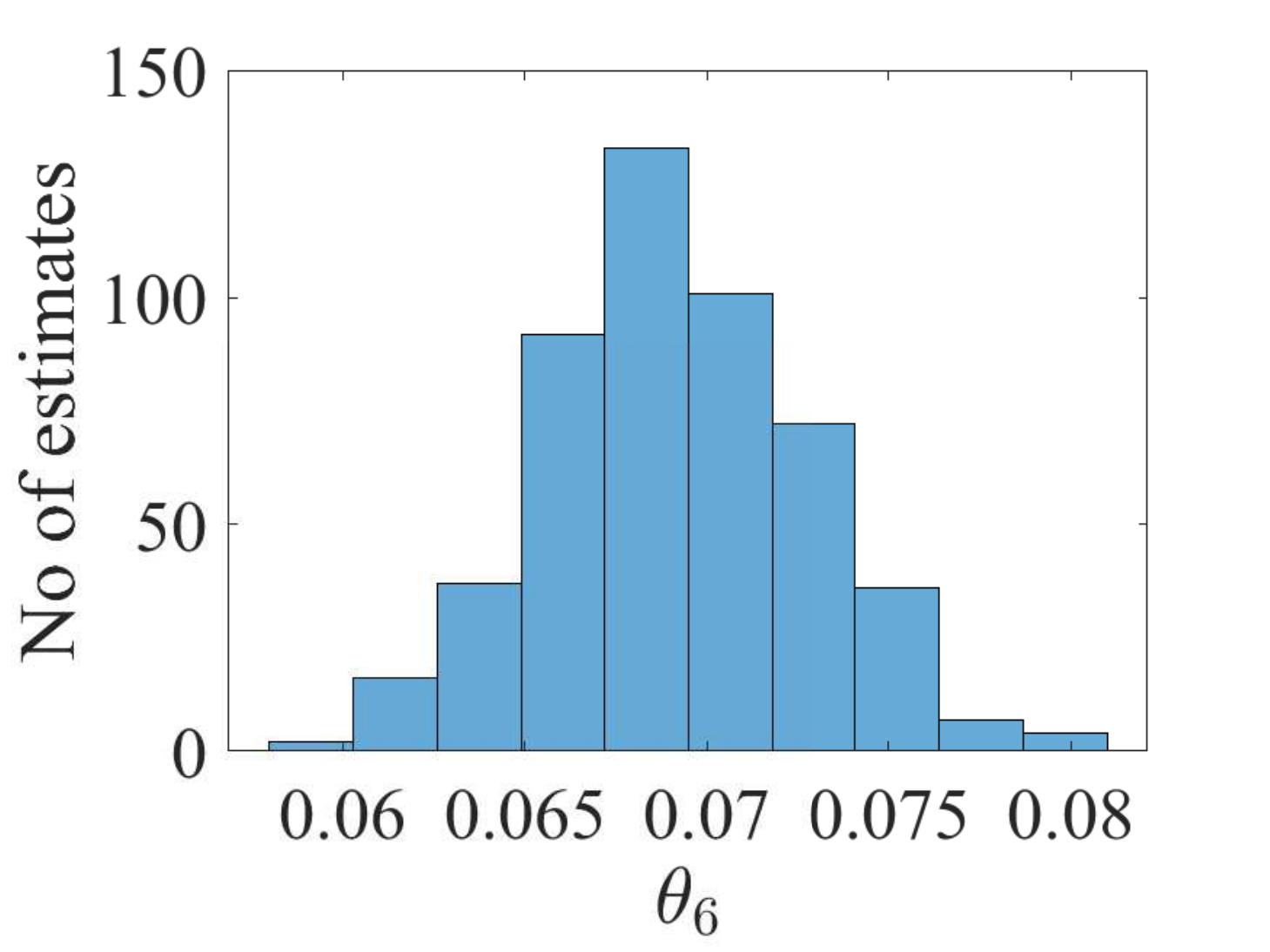}\label{Fig:Hist-RNLS}}
\centering
\caption{Histogram of estimates of $\theta_6$ for (a) NLS as a benchmark, (b) C-NLS and (c) R-NLS.} 
\label{Fig:3NLS-Hist}
\vspace{0mm}
\end{figure}

Figure~\ref{Fig:3NLS-Covariance} shows the NRMSE obtained by solving the original NLS problem~\eqref{Prediction-error-minimization} and applying the C-NLS and R-NLS methods, respectively. It is first seen that the NLS leads to extremely poor estimation accuracy, due to the intrinsic non-convexity as aforementioned. By comparison, both the C-NLS and R-NLS methods achieve substantial success, with the NRMSE lying below $10\%$ for every parameter. Going deeper, let us look in more detail at $\theta_6$, i.e., the resistance $\beta_1$. Figure~\ref{Fig:3NLS-Hist} shows the histogram  of the estimates of $\theta_6$ produced by the three methods. As is seen, the NLS estimation constantly strays far away from the truth, but the C-NLS and R-NLS give estimates near the true value 0.0678 with considerable consistency. These observations suggest that the proposed methods are highly competent in identifying the Thevenin's model parameters, even though the prior knowledge used is rough.

\begin{figure}[h]
\centering
{\includegraphics[trim = {0mm 0mm 0mm 0mm}, clip, width=0.4\textwidth]{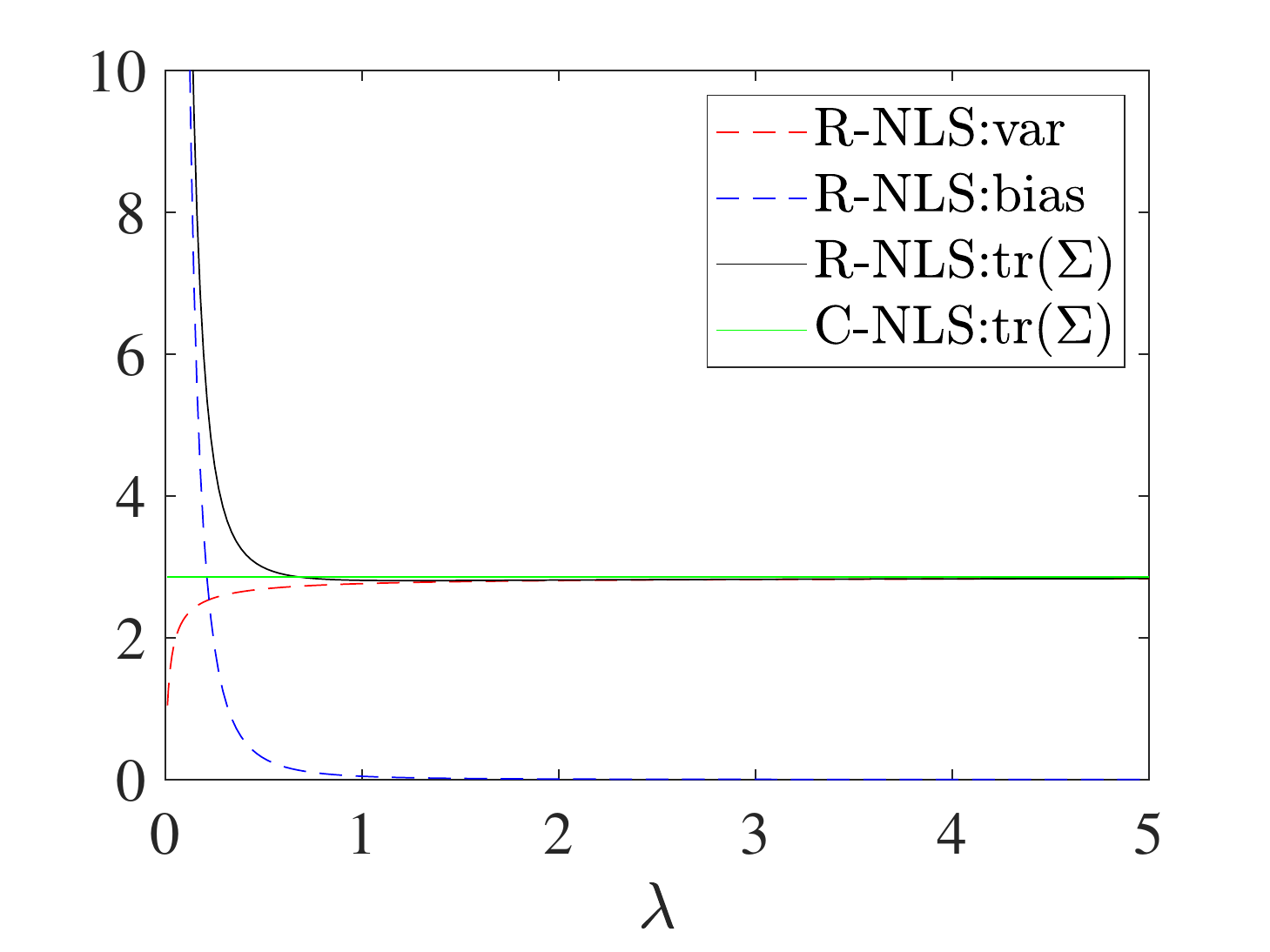}}
\caption{Theoretical estimation accuracy of the R-NLS method changes with $\lambda$.} 
\label{Fig:Lambda-Curve}
\vspace{0mm}
\end{figure}

Taking a closer look at Figure~\ref{Fig:3NLS-Covariance}, one can find out that  the estimation errors calculated based on the Monte Carlo simulation results match well with the theoretical values. This validates the correctness of the theoretical analysis. In addition, an implication specific for the R-NLS is that one can make use of~\eqref{RSigma-Accuracy} to select  $\bm\theta_0$ and $\bm{P}_0$. To explore this implication,  let us consider applying $\{\bm\theta_0,\lambda\bm{P}_0\}$ with $\lambda$ varying between 0 and 5 and $\{\bm\theta_0,\bm{P}_0\}$ shown in Table~\ref{Tab:Parameters-Setting}. Figure~\ref{Fig:Lambda-Curve} depicts the theoreical accuracy of the R-NLS with respect to $\lambda$, where var and bias correspond to the two terms in the rightmost-hand side of~\eqref{RSigma-Accuracy}, respectively. Several observations can be drawn. First, $\lambda$ plays a role in the estimation performance, or in other words, the selection of $\bm \theta_0$ and $\bm P_0$ will affect the eventual estimation accuracy. Second, the estimation accuracy will improve when $\lambda$ increases and then keep at a same level after $\lambda$ becomes large enough, and the nominal bias will approach zero. This indicates that the prior knowlege should be imposed with appropriate confidence and that it is wise to adopted a relatively large $\bm P_0$ if the prior knowledge is not precise. Finally, based on such a plot, a practitioner can choose $\bm \theta_0$ and $\bm P_0$ deemed as the best.


Concluding this section is a comparison of the running time. All the  simulations in this section are conducted on a Dell Precision Tower 3620 with 3 GHz Inter Xeon CPU and 16 Gb RAM  running MATLAB R2018b. The computation time is averaged on the results of the 500 Monte Carlo runs. The original NLS,  C-NLS and  R-NLS on average take 88 ms, 142 ms, and 84 ms, respectively. From this result, one can find that  both the C-NLS and R-NLS methods are computationally well affordable, especially given that they are designed for offline  identification. Furthermore, the R-NLS computes faster, which comes as an additional advantage, because the regularization expedites the parameter search.  

\section{Experimental Validation}\label{Sec:Experimental-Validation}
\begin{figure}[t]
\centering
\includegraphics[trim = {0mm -15mm 0mm 0mm}, clip, width=0.38\textwidth]{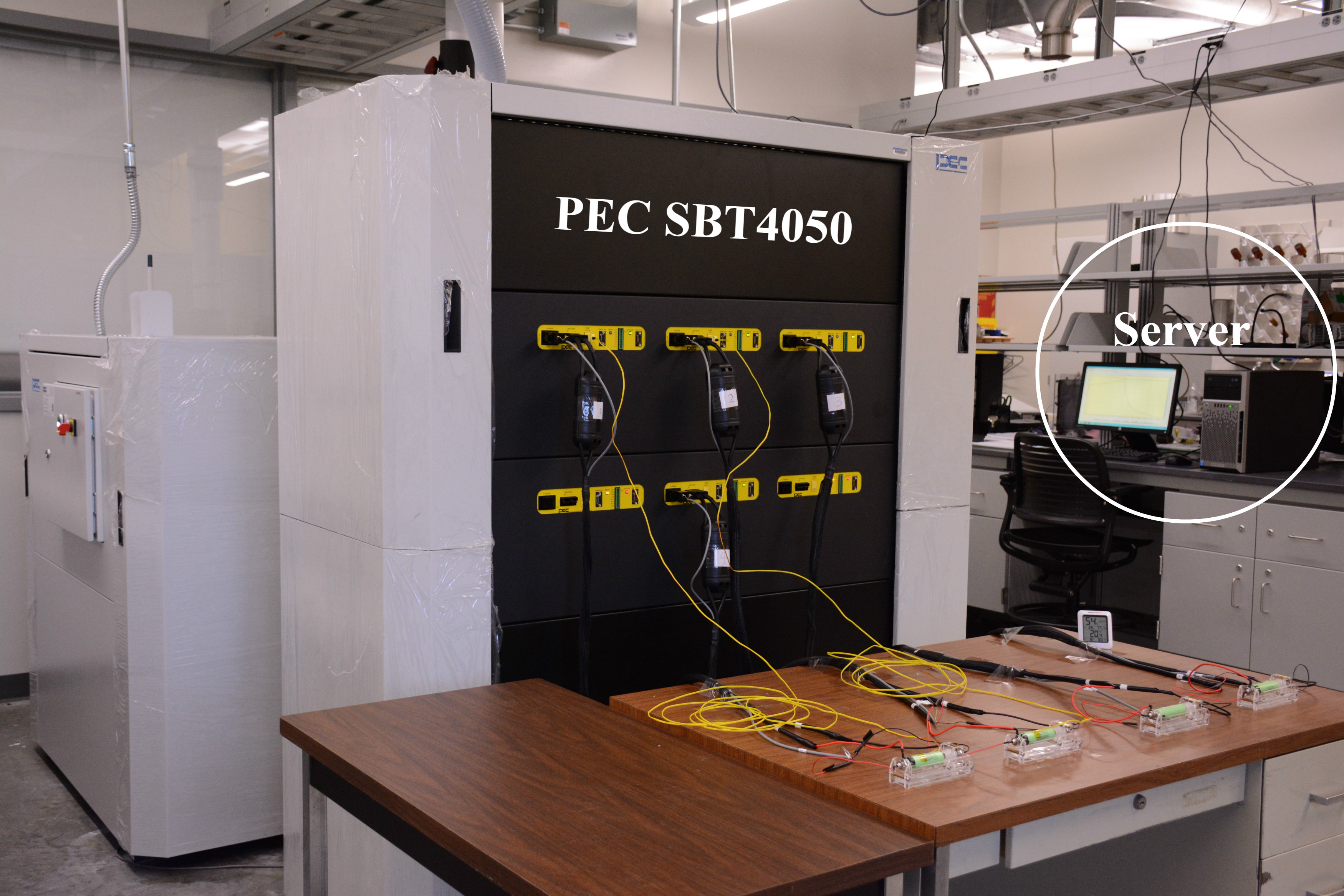}
\caption{The experimental facility.}
\label{Fig:Experiment-System}
\vspace{0mm}
\end{figure}

This section validates the C-NLS and R-NLS methods on experimental data. The experiments were conducted using a PEC\textsuperscript{\textregistered}
SBT4050 battery tester, which is shown in Figure~\ref{Fig:Experiment-System}. This tester has six channels and supports charging/discharging with arbitrary current-, voltage-, and power-based loads (up to 40 V and 50 A). It runs with a specialized server, which is used to   configure a test offline and collect experimental data online, both through the associated software application LifeTest\textsuperscript{TM}.  The experiments were performed on a Samsung INR18650-25R Li-ion battery. One test was conducted first to generate the training data, from which the parameters are extracted. Then, three more experiments were performed to generate validation datasets, to which the identified model is applied to assess its predictive performance. 

\subsection{Parameter Identification with Training Data}
\begin{figure}[t]
\centering
\includegraphics[trim = {2mm 5mm 2mm 14mm}, clip, width=0.83\textwidth]{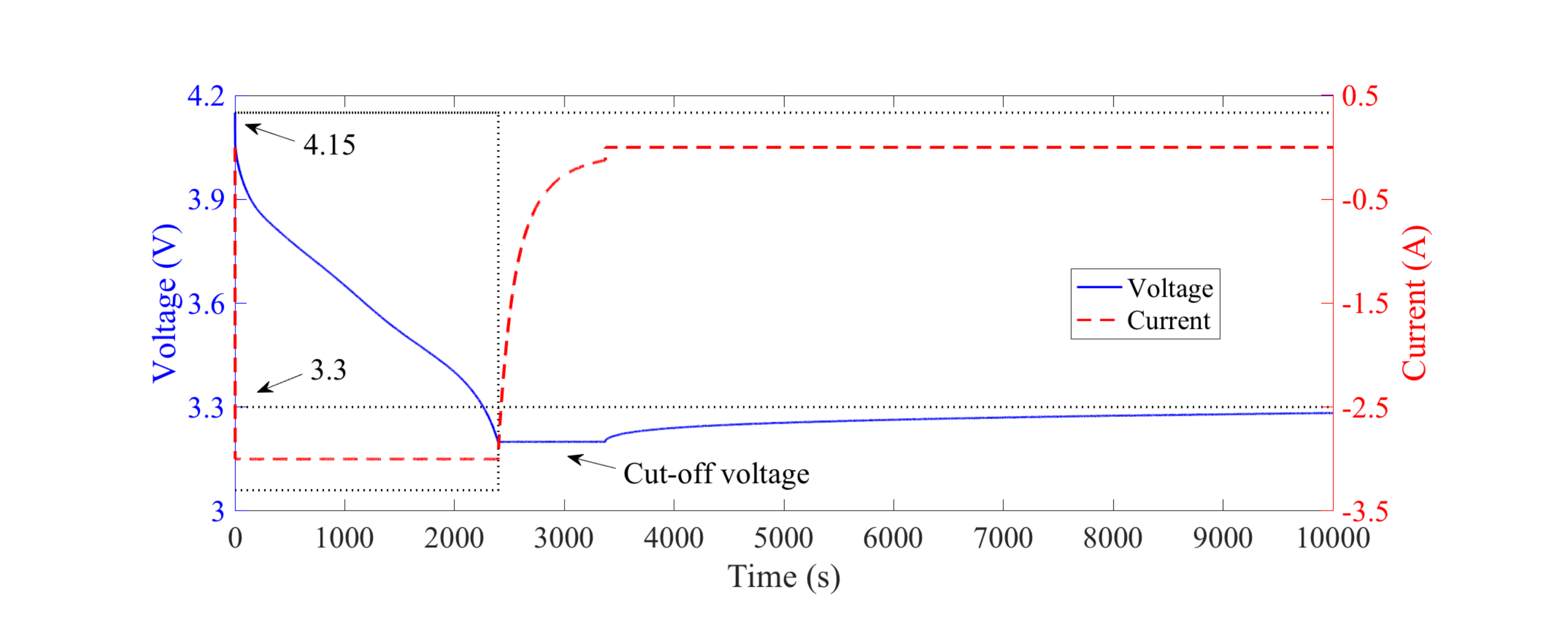}
\caption{The voltage response of the Samsung  INR18650-25R  Li-ion battery  in a constant-current/constant-voltage discharging experiment.}
\label{Fig:Cycle-Discharge}
\vspace{0mm}
\end{figure}

Considering $\underline{V}_{\rm OC}=3.3~\rm V$ and $\overline{V}_{\rm OC}=4.15~\rm V$, the first experiment   discharged the battery under room temperature using a constant current of -3 A from full charge until the terminal voltage hit the cut-off threshold of 3.2 V and then switched to constant-voltage discharging until the battery was fully depleted (when the current reduced to 0.125 A).  The voltage asymptotically recovered to $3.3~\rm V$ after the battery rested for a long time.  Figure~\ref{Fig:Cycle-Discharge}   shows the obtained current and voltage profiles. It should be noticed that  a constant-voltage discharging phase was included here in order to determine the battery's total capacity. One can safely remove it if the capacity is known prior. Based on Figure~\ref{Fig:Cycle-Discharge}, the battery's nominal capacity was calculated as $Q_{\rm c}=2.17~\rm Ah$ using coulomb counting~\cite{lam2011practical}.  Provided the $I$-$V$ data in the constant-current discharging stage  (encompassed by the black dotted-line box in Figure~\ref{Fig:Cycle-Discharge}),  the C-NLS and R-NLS methods were applied to extract parameter estimates, using the information setting  in Table~\ref{Tab:Parameters-Setting}. Meanwhile, the original NLS problem~\eqref{Prediction-error-minimization} was also solved as a benchmark.

Before  examining the parameter identification results, let us explain how the information  in Table~\ref{Tab:Parameters-Setting} is determined through the summary below. Here, the initial guess is used to initialize the  identification run for all the methods;   $\underline{\bm \theta}$ and $\overline{\bm \theta}$ are the  parameter bounds for the C-NLS, and  $\bm \theta_0$ and $\bm{P}_0$ represent the advance knowledge used by the R-NLS.
\begin{itemize}
\item The parameters $\alpha_1$ through $\alpha_4$ are hard to guess  in advance as   they have no physical meaning, so they are  simply initialized to be one. There is also no need to assign  bounds for them  in the C-NLS case. In the R-NLS case, it is  safe to leave the elements of $\bm\theta_{0}$ and $\bm{P}_0$ corresponding to them as blank,  but if  one wants to calculate the theoretical accuracy using~\eqref{RSigma-Accuracy}, it is acceptable to set  $\bm\theta_{0}$ without much care about accuracy and meanwhile set a large $\bm{P}_0$   as in Table~\ref{Tab:Parameters-Setting}.

\item The parameter $\beta_0$ is roughly equal to the immediate voltage drop upon the start of discharging divided by the applied current. From Figure~\ref{Fig:Cycle-Discharge}, it is estimated as 0.029. Since this guess can be close to the truth, it is reasonable to stipulate a prior $99\%$ confidence interval of  $0.029\pm0.003$  for $\beta_0$, where the standard deviation is $0.001$, to run the R-NLS.   For the execution of the C-NLS, the bounds are loosely set as 0.01 and 0.04. 

\item The voltage recovery after the end of the entire discharging process is largely due to the loss of voltage across $R_0({\rm SoC}=0)=\beta_0+\beta_1$. Hence, $\beta_1$ is upper bounded by $0.1/0.125=0.8$ based on Figure~\ref{Fig:Cycle-Discharge}.   Its lower bound is assigned to be zero for simplicity. The average of the bounds, 0.4, hence is chosen as the initial guess of $\beta_1$, along with    a standard deviation of $0.4/3$.

\item The parameter $\beta_2$  mostly determines the curve shape of $R_0({\rm SoC})$. Its lower and upper bounds are very loosely set as 0 and 80. 
It is hence initially guessed as 40, which is the average of the bounds, with a standard deviation of  $40/3$. 

\item The total voltage decline, following the immediate voltage drop when the discharging starts and lasting until when the constant-current discharging terminates, is from 4.15 V to 3.2 V. This is contributed   by  the combined change in the OCV and the voltage across $R_0$ and $R$. Note that the voltage across $R$ will also approach a constant (i.e., $IR$) after sufficiently long time. The information can be used to infer an upper bound for $R$ at 0.4 and a simple lower bound at 0. The average of the bounds, 0.2,  is set as the initial guess,  and the standard deviation  set as 0.2/3.


\item Because $RC$ is the time constant for the RC circuit, its value can be roughly seen from the voltage curve during the constant-current discharging. The voltage starts to decline almost linearly after the first 200 s,   implying that $RC$ is vaguely around $200/5=40$. The initial guess of $(RC)^{-1}$ is thus taken to be $1/40$, with a standard deviation of $0.005$. Loose lower and upper bounds are chosen, which are 1/200 and 1, respectively.

\end{itemize}

From above, one can develop some advance knowledge of the parameters through straightforward observation and analysis of the data and build an information setting as shown in Table~\ref{Tab:Parameters-Setting}. This allows the proposed C-NLS and R-NLS methods to easily lend themselves to use in practice.

\begin{table}[t]\centering
\begin{threeparttable}
\caption{Parameter estimates for the battery used in the experiments.}
\begin{tabular}{ccccccccccc}
\toprule%
Name& $\alpha_1/1$ &$\alpha_2/1$ &$\alpha_3/1$ & $\alpha_4/1$  & $\beta_0/\Omega$    & $\beta_1/\Omega$    & $\beta_2/1$ & $R/\Omega$   & $(RC)^{-1}/{\rm s^{-1}}$\\
\midrule
$\hat{\bm \theta}_{\rm Benchmark}$ &  3.48 & -11.3 & 20.5  &-17.0  &0.0360  &0.0465 & 155  & 0.0668   & 0.0070  \\ 
$\hat{\bm \theta}_{\rm C-NLS}$ &  2.61 & -9.36 & 19.7  & -19.0  &0.0313 &0.0678 &   13.2 & 0.0313   & 0.0172  \\ 
$\hat{\bm \theta}_{\rm R-NLS}$ &  2.60 & -9.34  & 19.7 & -19.0   &0.0308 &0.0689 &  13.2 &  0.0313 & 0.0179   \\
\bottomrule
\end{tabular}
\label{Tab:Identification-Results-Experiment}
\end{threeparttable}
\end{table}

\begin{figure}[t]
\centering
\subfigure[]
{\includegraphics[trim = {0mm 0mm 0mm 0mm}, clip, width=0.3\textwidth]{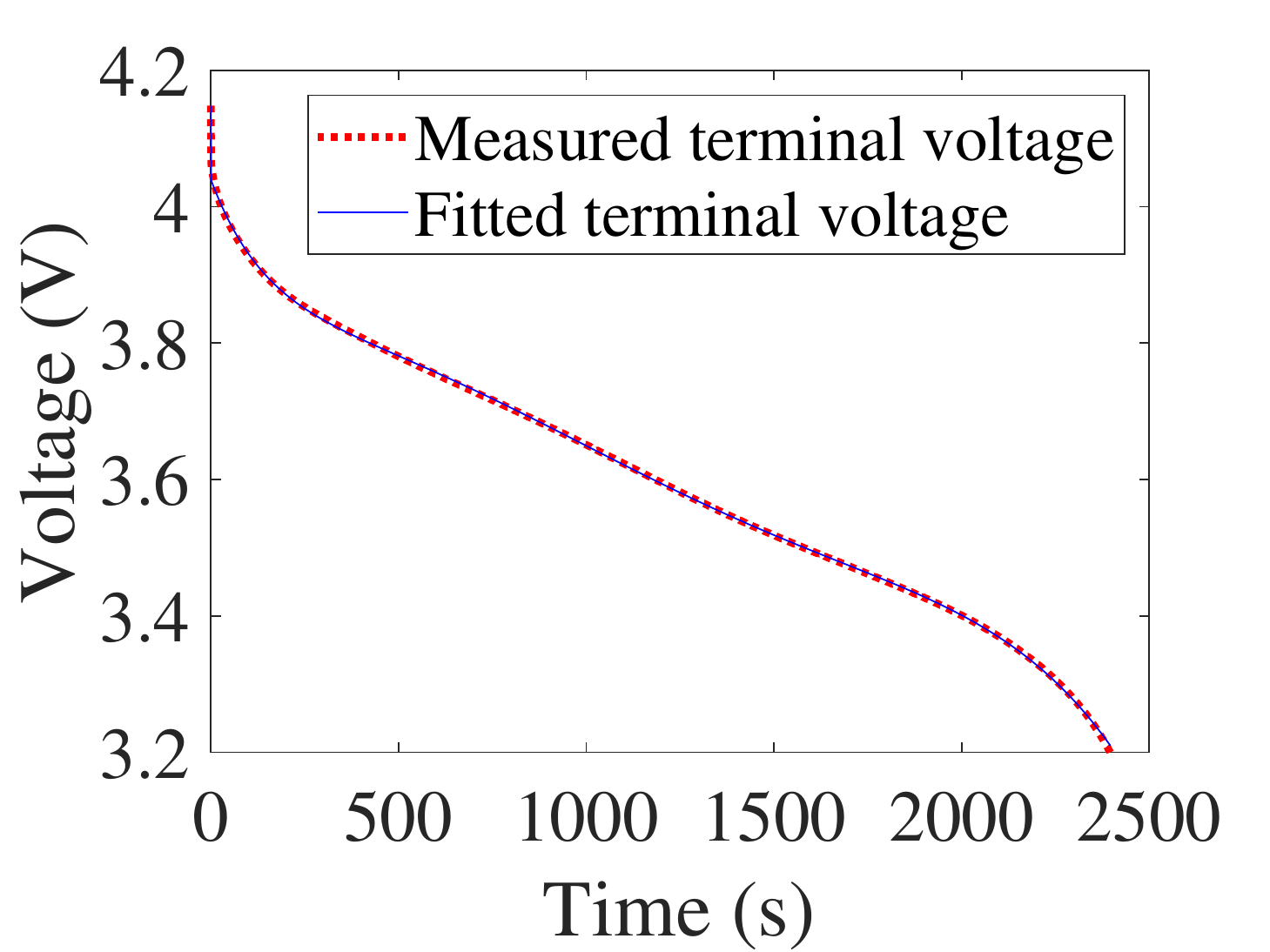}\label{Fig:Fitting-Curves-Experiment-Voltage-a}}
\subfigure[]
{\includegraphics[trim = {0mm 0mm 0mm 0mm}, clip, width=0.3\textwidth]{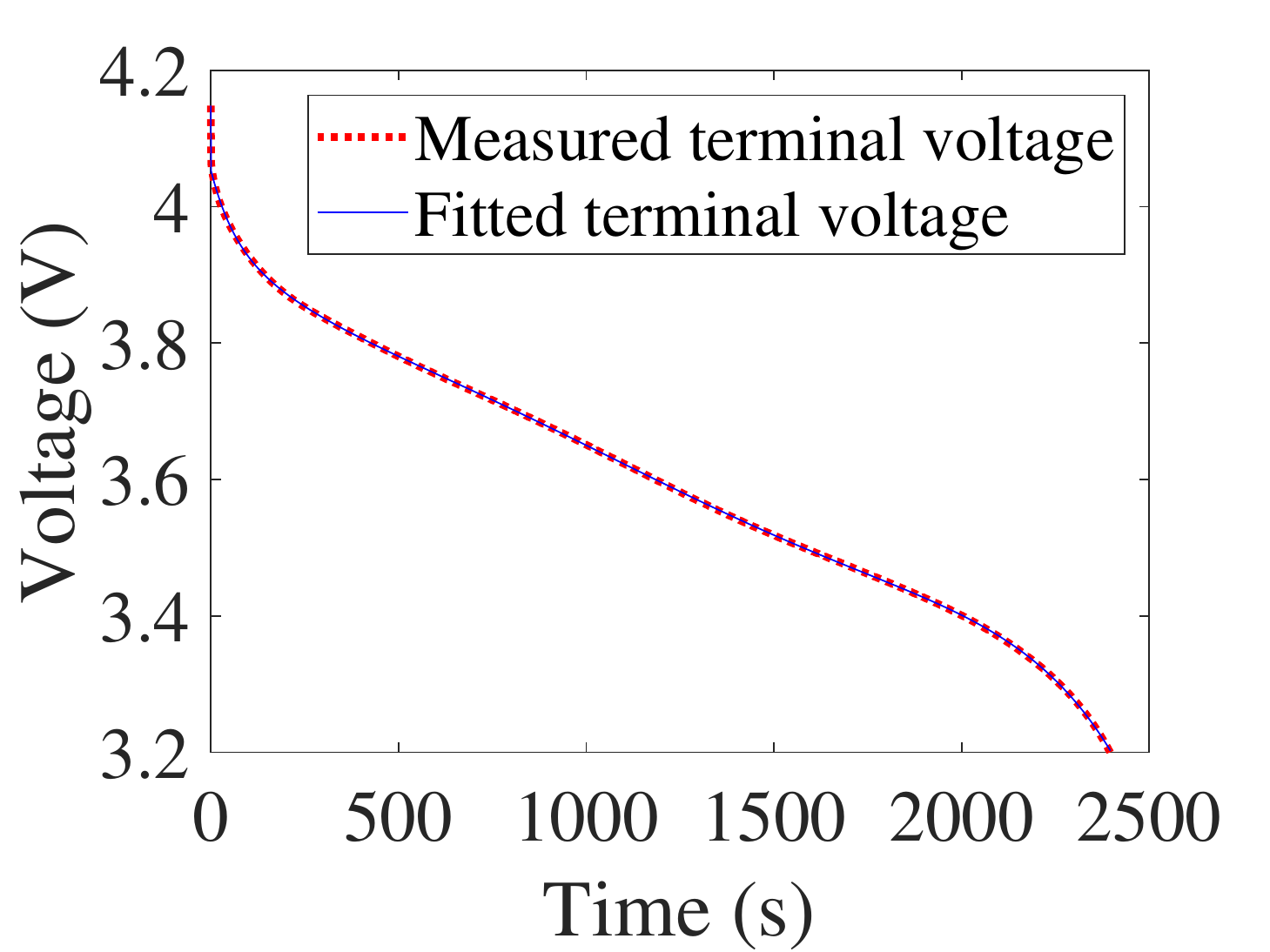}\label{Fig:Fitting-Curves-Experiment-Voltage-b}}
\subfigure[]
{\includegraphics[trim = {0mm 0mm 0mm 0mm}, clip, width=0.3\textwidth]{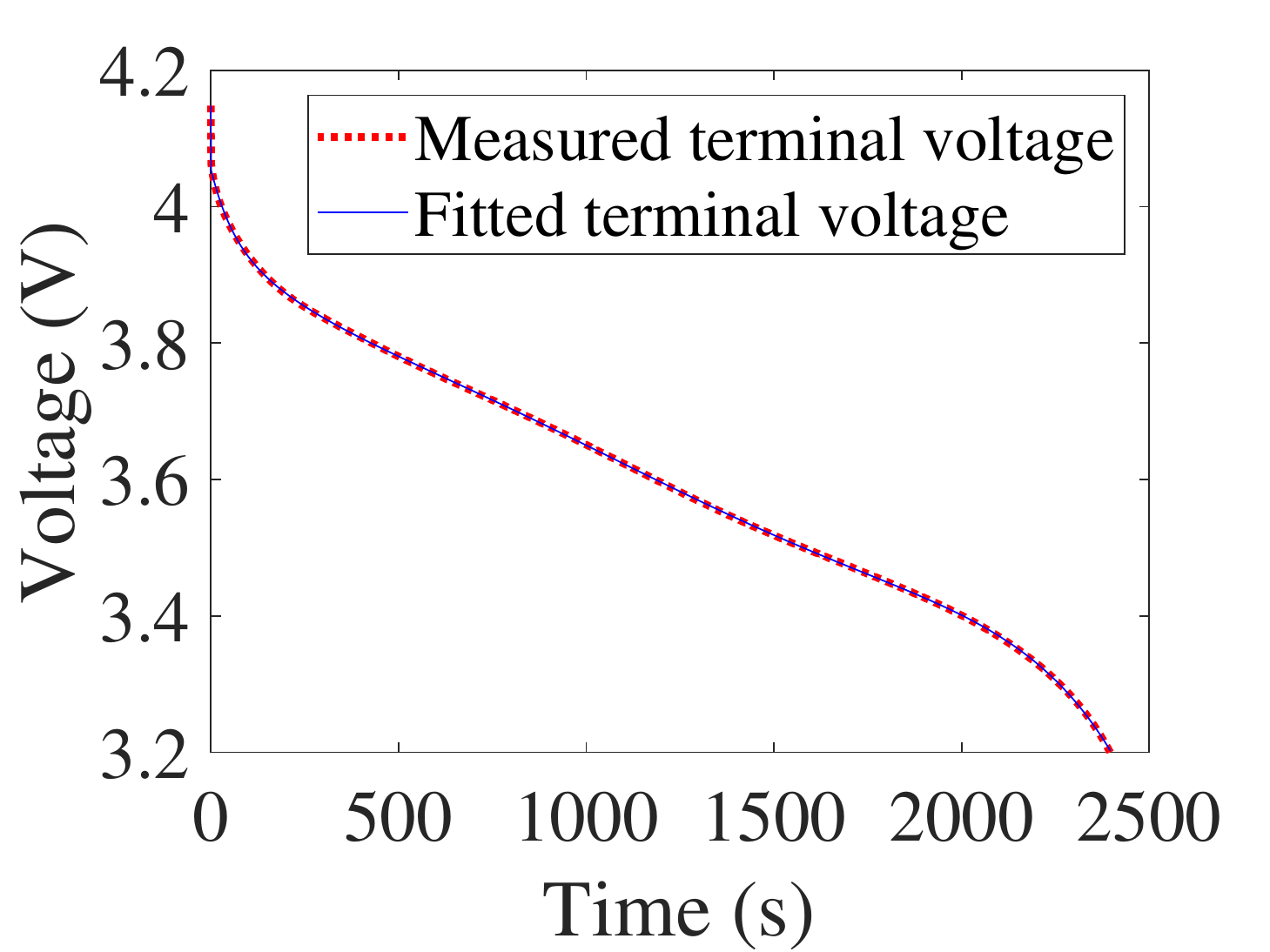}\label{Fig:Fitting-Curves-Experiment-Voltage-c}}
\centering
\caption{Comparison of measured and fitted terminal voltage for (a) NLS as a benchmark, (b) C-NLS and (c) R-NLS. }
\label{Fig:Fitting-Curves-Experiment-Lab}
\vspace{0mm}
\end{figure}


With the above   settings and letting $\bm Q=2.5\times10^{-5}\bm I$, the original NLS, C-NLS and R-NLS are applied to estimate the parameters. The parameter estimates are summarized in Table~\ref{Tab:Identification-Results-Experiment}, and the voltage data fitting  shown in Figure~\ref{Fig:Fitting-Curves-Experiment-Lab}. Figure~\ref{Fig:Fitting-Curves-Experiment-Lab} displays that  the predicted terminal voltage  well overlaps the measurements in all   three cases.  However, with a look at Table~\ref{Tab:Identification-Results-Experiment}, it is interesting to find out that the C-NLS and R-NLS produce very close estimation, whereas the estimates of the  original NLS   are quite different. Next, validation data are utilized to further evaluate the quality of the three identified models.

\begin{figure}[t]
\centering
\includegraphics[trim = {21mm 21mm 17mm 47mm}, clip, width=0.9\textwidth]{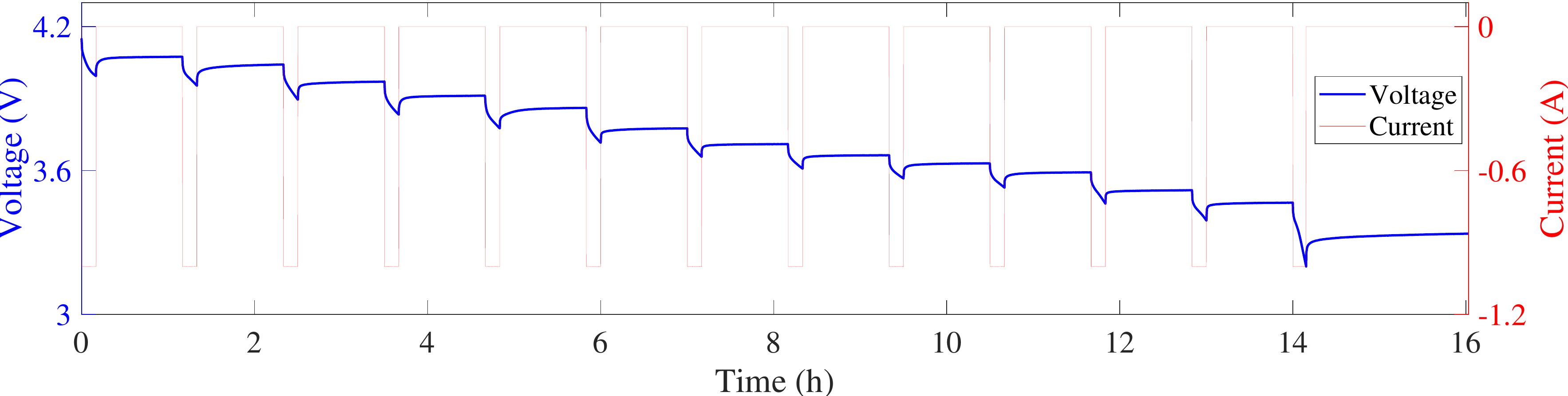}
\centering
\caption{The voltage response of the battery in an intermittent discharging test.} 
\label{Fig:SoC-OCV-Experiment}
\vspace{0mm}
\end{figure}

\begin{figure}[t]
\centering
\subfigure[]
{\includegraphics[trim = {0mm 0mm 0mm 0mm}, clip, width=0.3\textwidth]{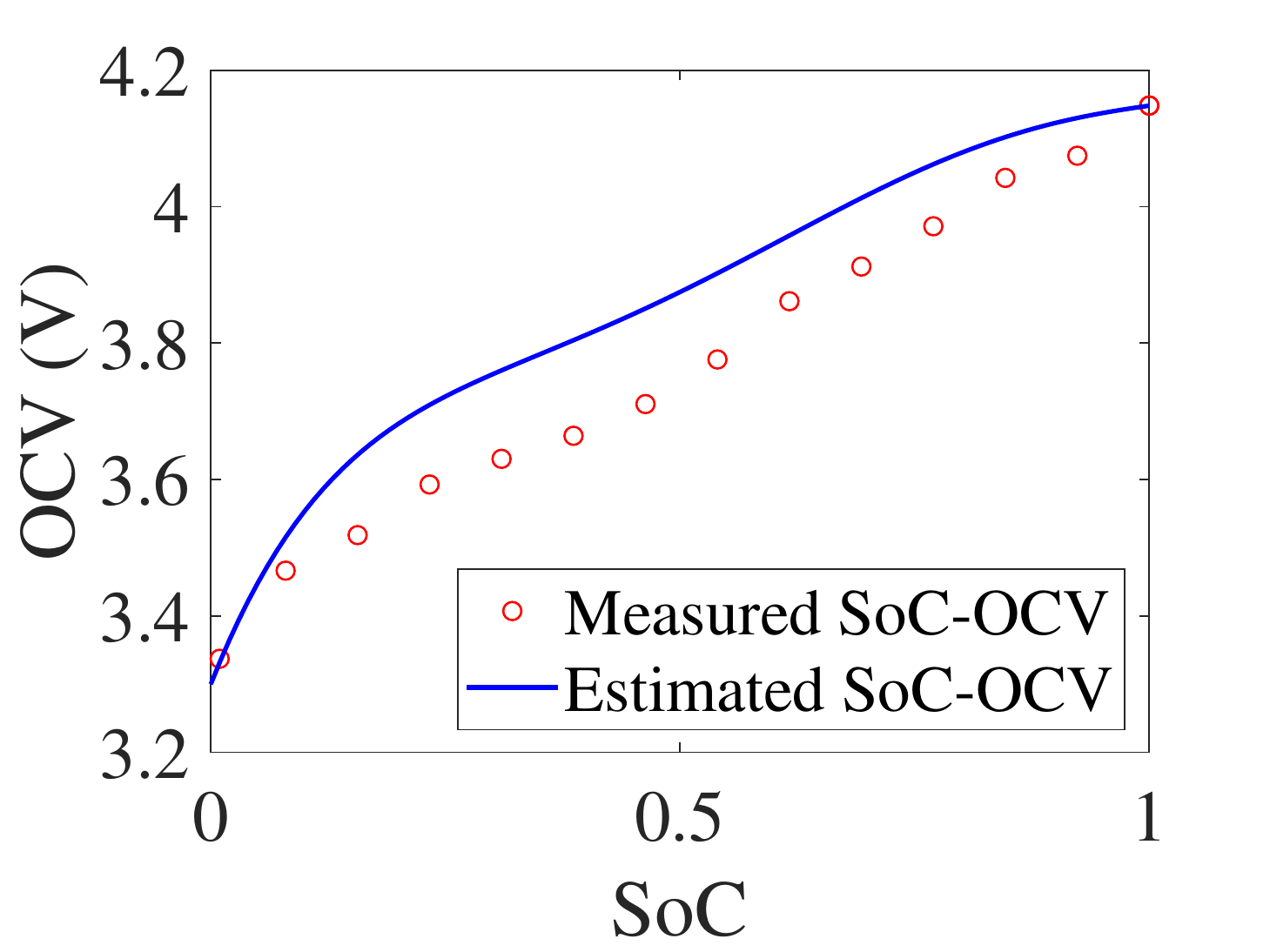}\label{Fig:SoC-OCV-Validation-N}} 
\subfigure[]
{\includegraphics[trim = {0mm 0mm 0mm 0mm}, clip, width=0.3\textwidth]{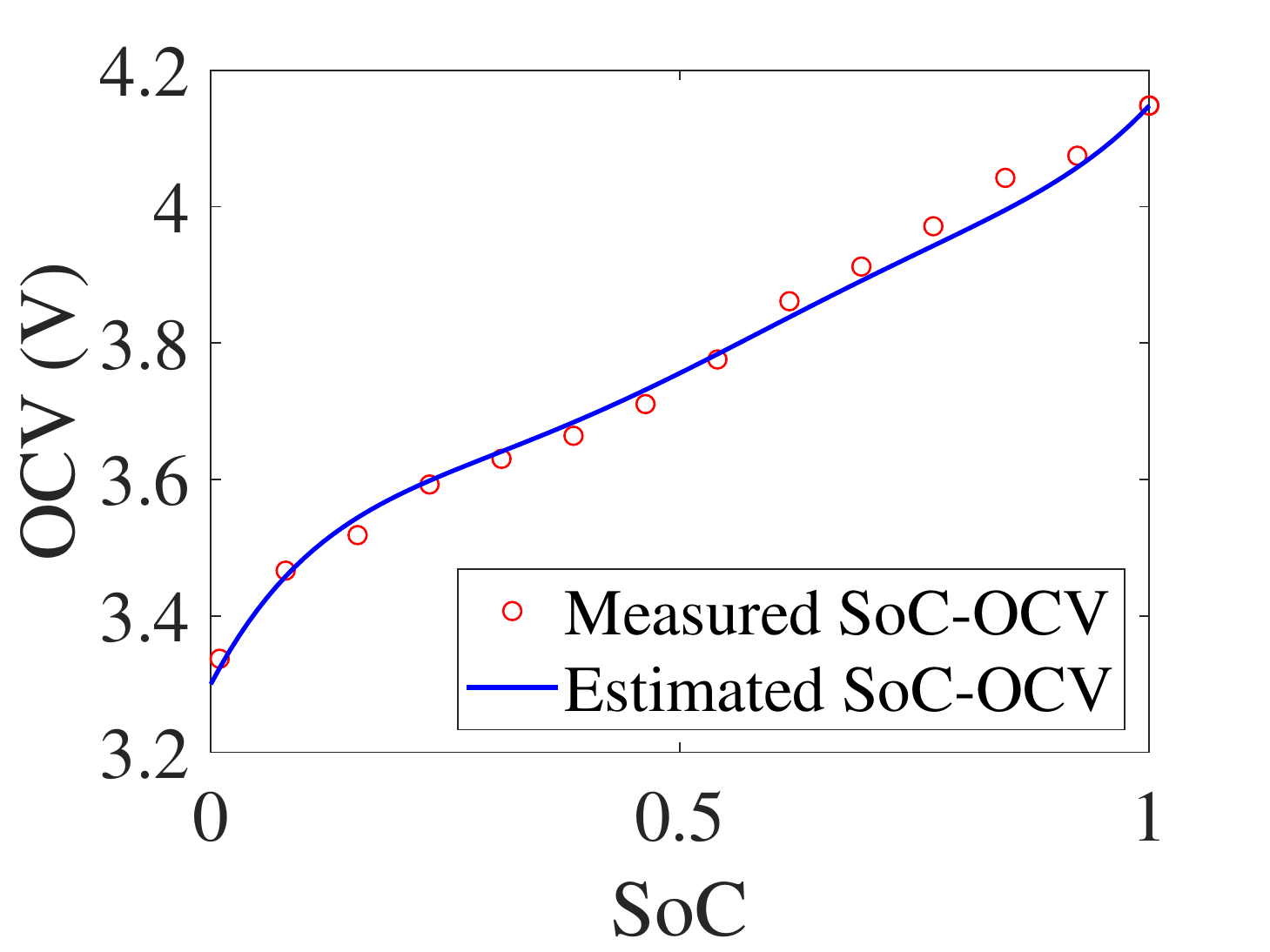}\label{Fig:SoC-OCV-Validation-C}}
\subfigure[]
{\includegraphics[trim = {0mm 0mm 0mm 0mm}, clip, width=0.3\textwidth]{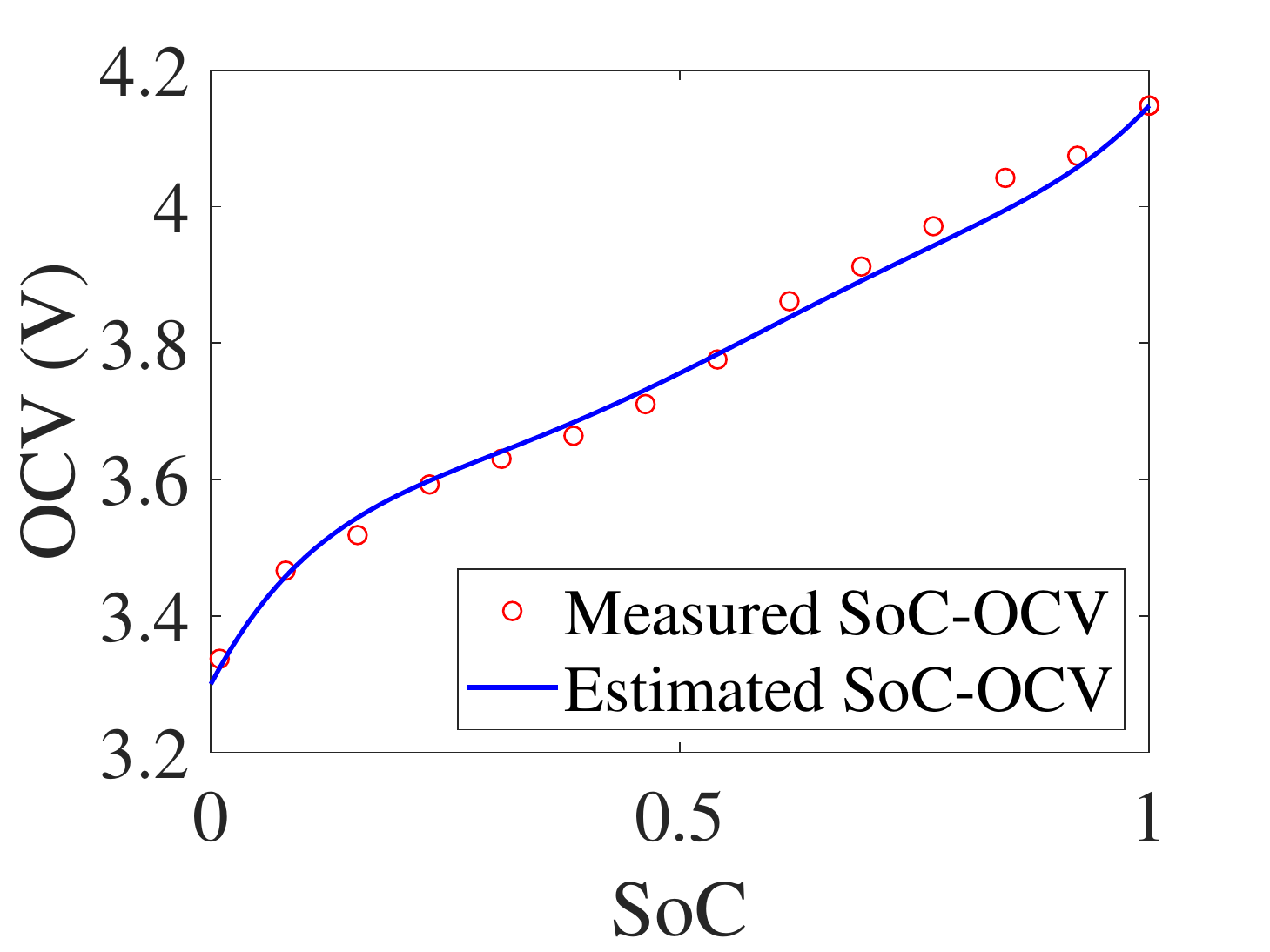}\label{Fig:SoC-OCV-Validation-R}}
\centering
\caption{Comparison of measured and estimated SoC-OCV for (a) NLS as a benchmark, (b) C-NLS and (c) R-NLS.}
\label{Fig:SoC-OCV-Validation}
\vspace{0mm}
\end{figure}

\begin{figure}[!htbp]
\centering
\subfigure[]
{\includegraphics[trim = {0mm 0mm 0mm 0mm}, clip, width=0.3\textwidth]{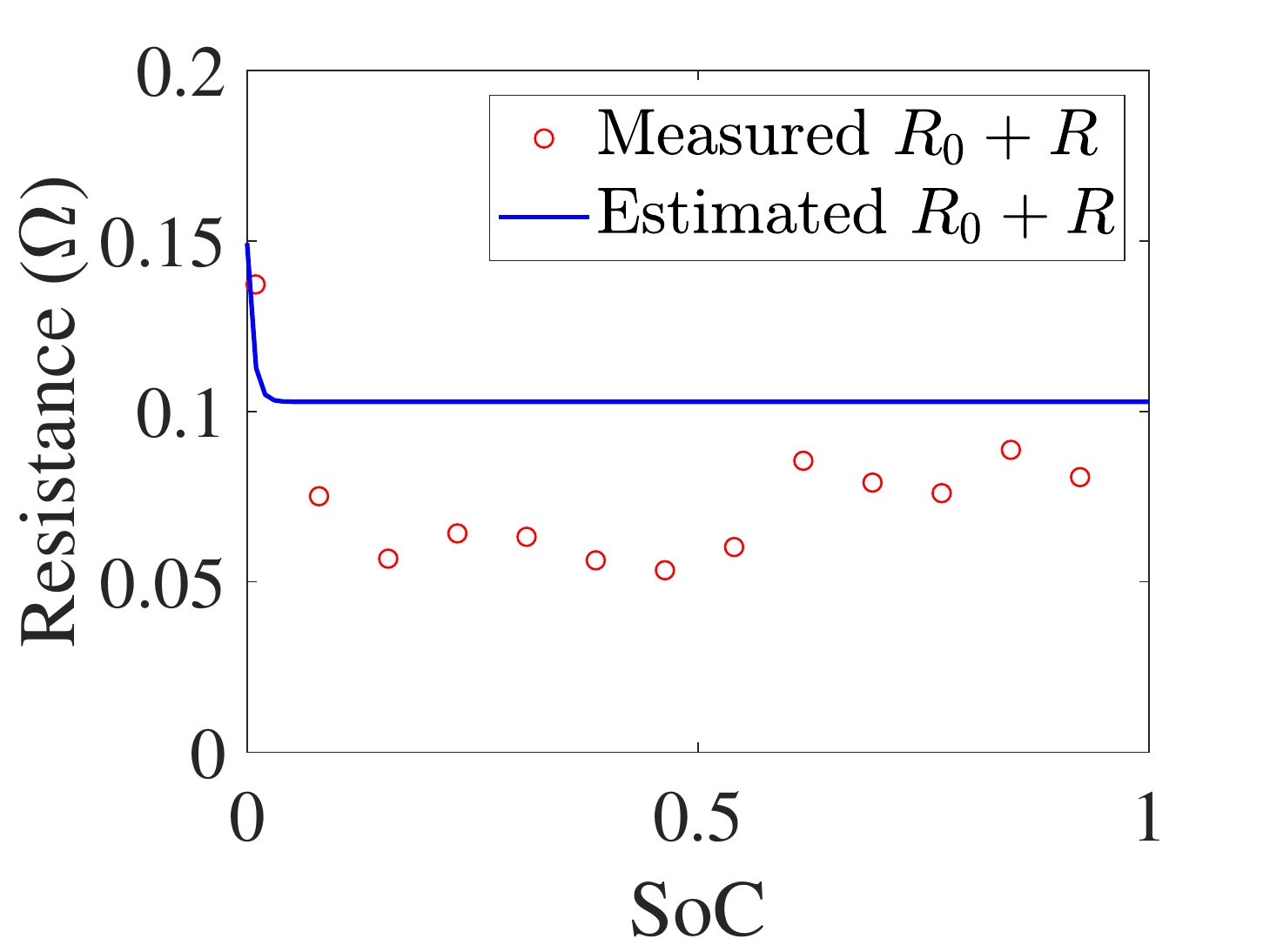}\label{Fig:R0R-Validation-N}} 
\subfigure[]
{\includegraphics[trim = {0mm 0mm 0mm 0mm}, clip, width=0.3\textwidth]{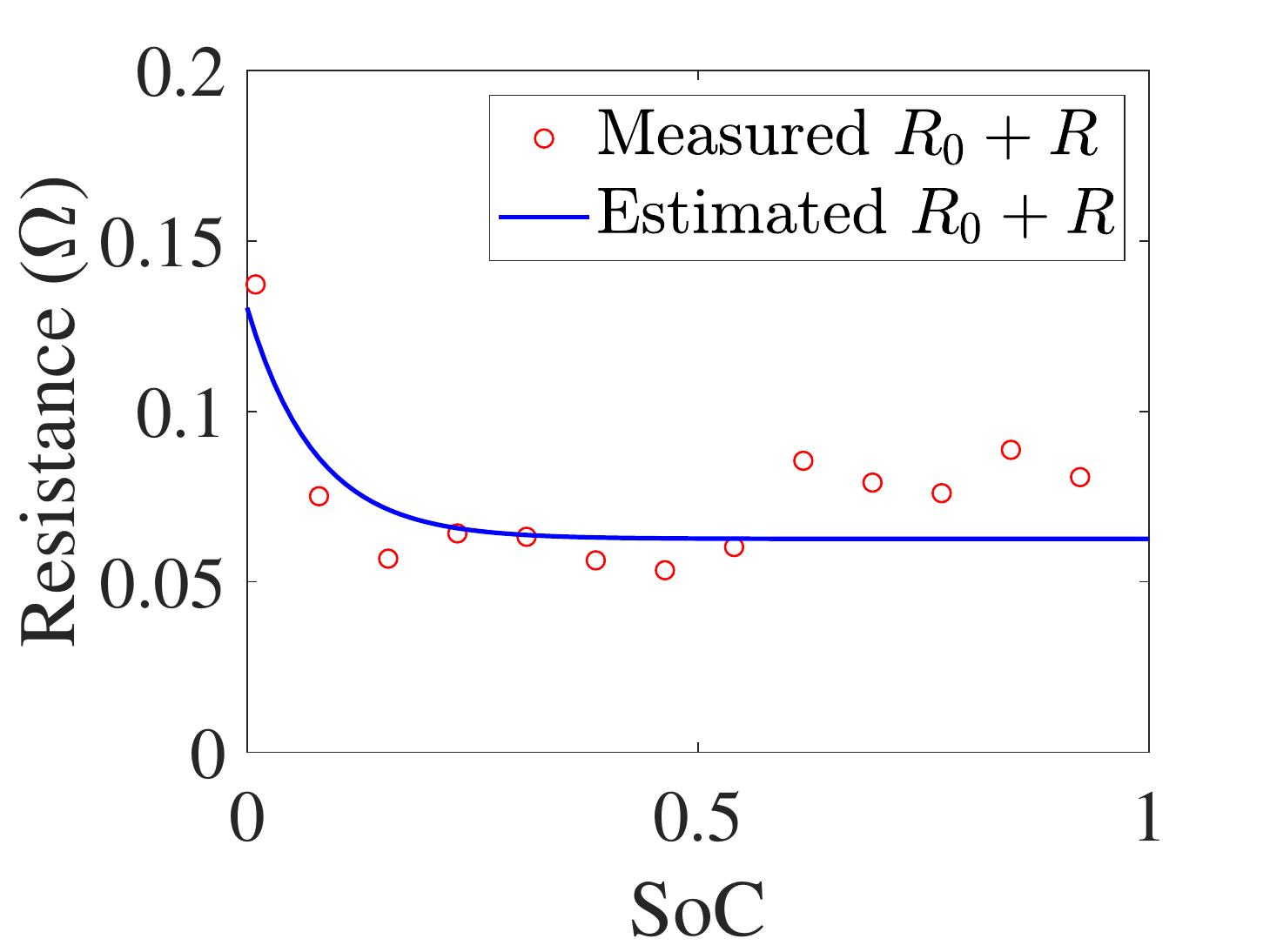}\label{Fig:R0R-Validation-C}}
\subfigure[]
{\includegraphics[trim = {0mm 0mm 0mm 0mm}, clip, width=0.3\textwidth]{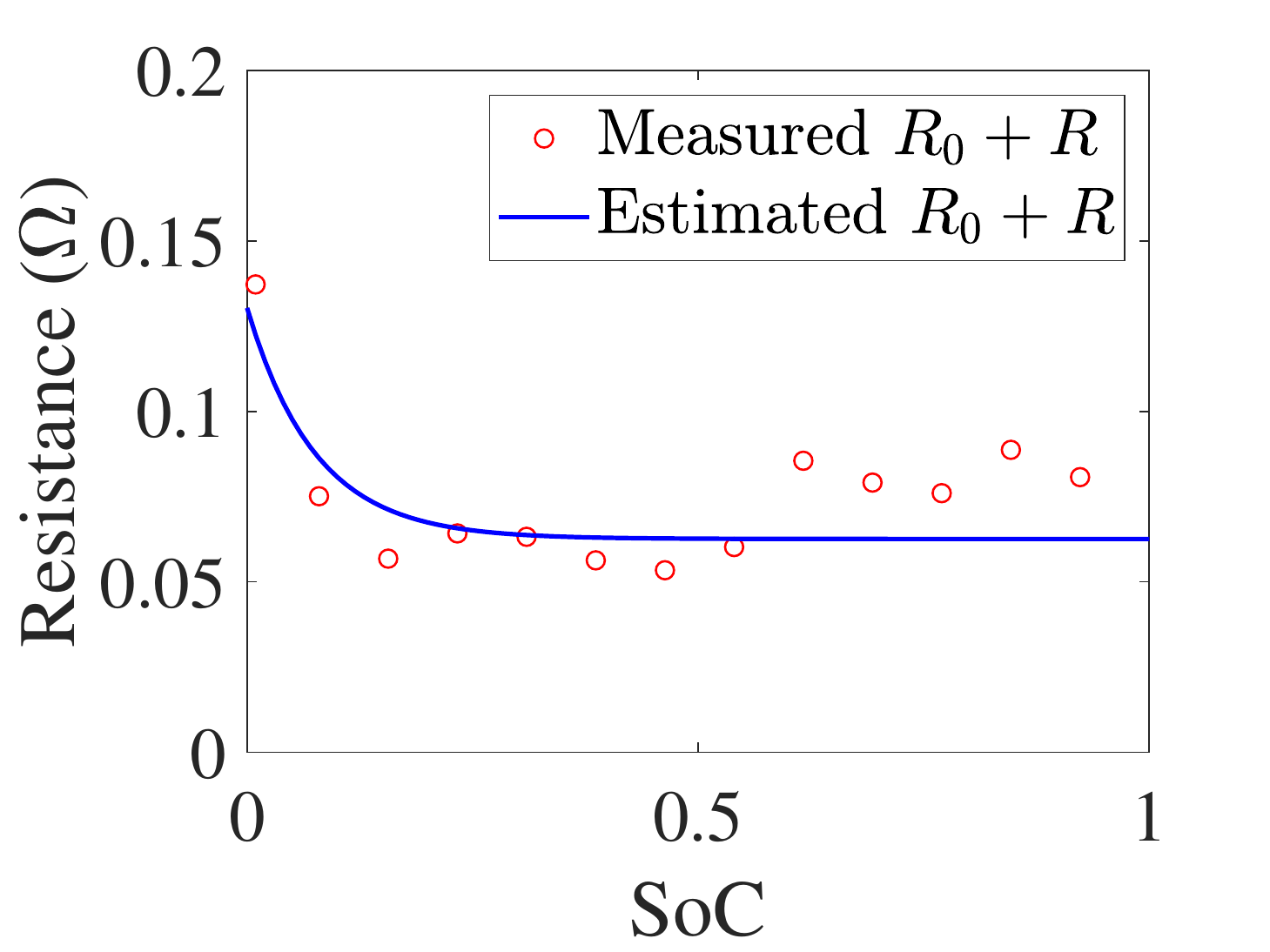}\label{Fig:R0R-Validation-R}}
\centering
\caption{Comparison of measured and estimated $R_0+R$ for (a) NLS as a benchmark, (b) C-NLS and (c) R-NLS.}
\label{Fig:R0R-Validation}
\vspace{0mm}
\end{figure}

\subsection{Model Validation with Validation Data}
Further experiments were conducted to generate datasets used to assess the performance of the above identified model. The first experiment, based on intermittent discharging, was designed to validate the accuracy of the estimated SoC-OCV relationships and impedance parameters. In this experiment, the battery was discharged by a constant current of $-1$ A for ten minutes and then left at rest for two hours such that the battery voltage would recover to a steady state. Such a  procedure repeated itself until the voltage declined to the cut-off voltage of 3.2 V. The current and voltage profiles of this experiment are shown in Figure~\ref{Fig:SoC-OCV-Experiment}. This dataset allowed us to perform the following validation to verify the effectiveness of the proposed C-NLS and R-NLS methods.

\begin{itemize}
\item  First, since the terminal voltage after a two-hour recovery can be regarded as the OCV, one can plot the SoC-OCV and use it as the ground truth to evaluate the estimated SoC-OCV relationship. Figure~\ref{Fig:SoC-OCV-Validation} offers such a comparison.  Evidently, the SoC-OCV relationships identified by the C-NLS and R-NLS methods match well with the measured one. Contrasting this is a serious discrepancy in the benchmark case based on the original NLS. 

\item Second, the voltage recovery upon a pause of discharging can be largely attributed to the change in the voltage across $R_0$ and $R$, which is equal to $I(R_0+R)$. Hence, $R_0+R$ corresponding to different SoC levels can be determined and used to appraise the estimated values, as shown in Figure~\ref{Fig:R0R-Validation}. The C-NLS and R-NLS are also observed to achieve  more precise estimation than the original NLS in this case.
\end{itemize}

\begin{figure}[t]
\centering
\includegraphics[trim = {30mm 0mm 40mm 5mm}, clip, width=0.9\textwidth]{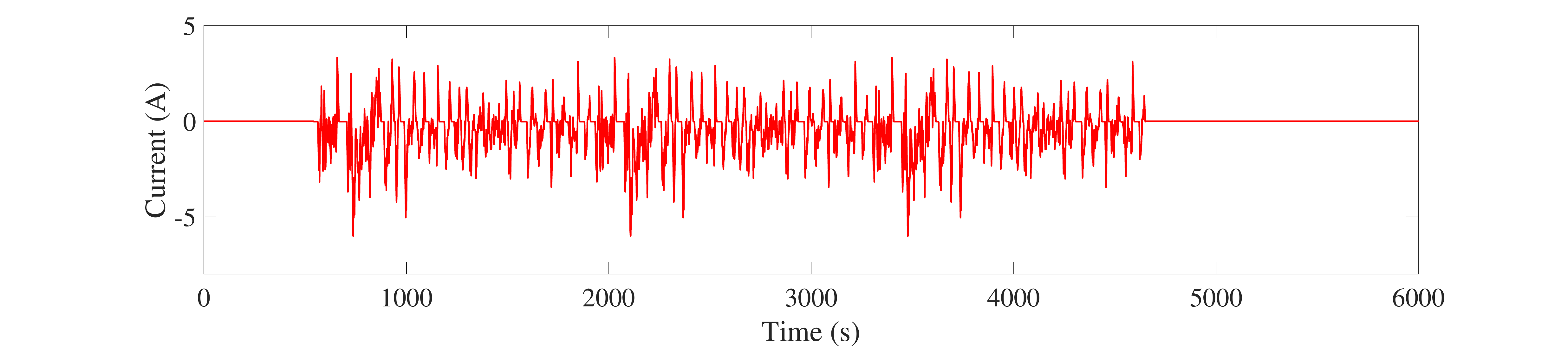}
\centering
\caption{Variable currents from normalization of the Urban Dynamometer Driving Schedule (UDDS) profile.} 
\label{Fig:Experiment-UDDS}
\vspace{0mm}
\end{figure}

\begin{figure}[t]
\centering
\subfigure[]
{\includegraphics[trim = {30mm 0mm 40mm 5mm}, clip, width=0.9\textwidth]{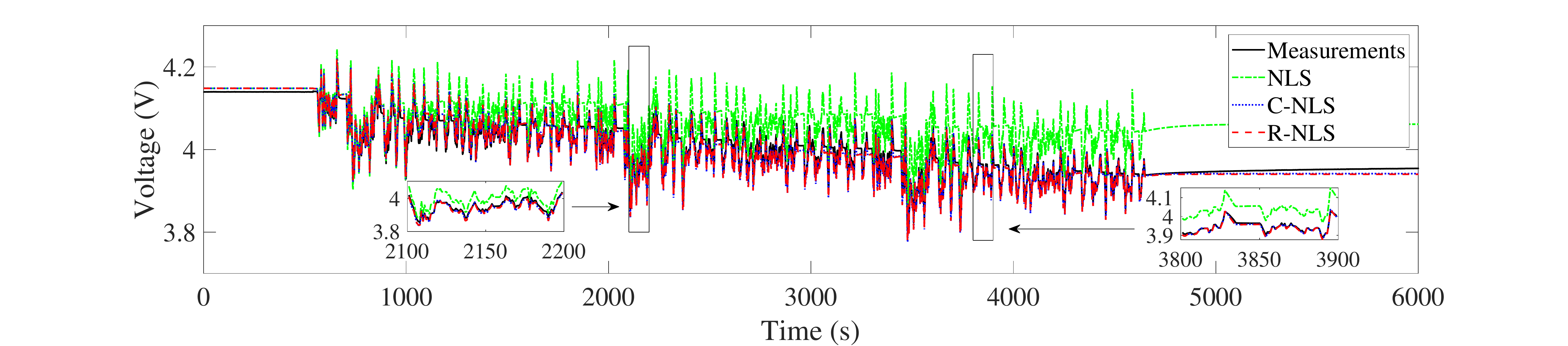}\label{Fig:UDDS-V-CR}}
\subfigure[]
{\includegraphics[trim = {30mm 0mm 40mm 5mm}, clip, width=0.9\textwidth]{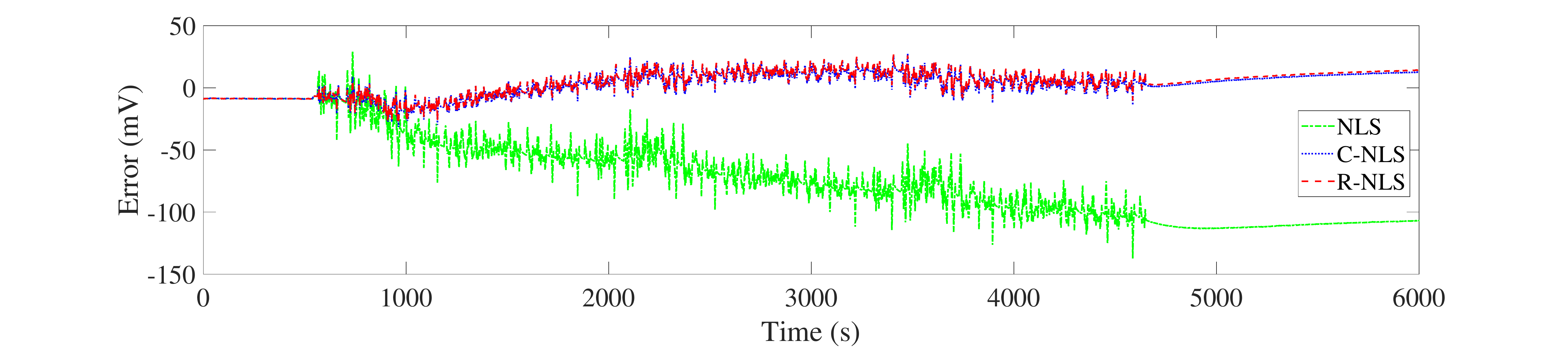}\label{Fig:UDDS-Error-CR}}
\caption{UDDS: (a) comparison of measured and estimated voltage for the benchmark NLS, C-NLS and R-NLS methods and (b) errors.} 
\label{Fig:UDDS-CR}
\vspace{0mm}
\end{figure}

Another experiment applied a variable current profile  to the battery, which was created based on    the Urban Dynamometer Driving Schedule (UDDS)~\cite{UDDS}.   Figure~\ref{Fig:Experiment-UDDS} displays the UDDS-based current profile, which involves both  charging and discharging in the run. Then, the identified models by the NLS, C-NLS and R-NLS methods are used to predict the terminal voltage under the   current profile. 
Figure~\ref{Fig:UDDS-V-CR} compares the predicted voltage  against the actual measurements. An overall excellent fitting accuracy can be observed for the models identified by the C-NLS and R-NLS, whereas for the benchmark NLS the accuracy is far from satisfactory.  Figure~\ref{Fig:UDDS-Error-CR} further plots the prediction errors, which are found to be  generally less than 20 mV for the C-NLS and R-NLS and larger by a great margin for the original NLS.  This validation provides clear-cut evidence that the identified models by the C-NLS and R-NLS are accurate,   justifying the effectiveness and competence of the two methods for parameter identification.  An experiment based on the  Worldwide harmonized Light-duty vehicles Test Procedure (WLTP)~\cite{WLTC} also shows similar results, though they are not reported here for the sake of space.

\section{Conclusion}\label{Sec:Conclusion}
This paper studied the open problem of one-shot parameter identification for the popular Thevenin's battery model. The objective was to build an approach capable of estimating all  model parameters  offline  from the current/voltage data  in a single run.  The study   formulated the identification problem in a prediction-error-based framework and conducted the identifiability analysis, unveiling that the parameters of the Thevenin's model are locally identifiable. Given the nonlinearity of  the identification problem, numerical optimization was considered, but  the non-convexity of the cost function   inflicted local minimum trouble on the optimization procedure. This work then took two approaches to deal with this challenge, one imposing parameter bounds to constrain the search space, and the other regularizing the cost function with prior knowledge of the parameters.  The resultant two identification methods, along with the identifiability analysis, were extensively validated through simulation and experiments. The proposed methods can provide high-quality parameter estimation and are relatively easy to implement with a one-shot run, having a  potential for  many battery management applications relying on accurate models, e.g., optimal charging, SoC estimation and monitoring of aging condition. Our future work will extend  the investigation to the identification of     Thevenin's models with temperature- and SoC-dependent parameters.

\appendix
\setcounter{secnumdepth}{0}
\section{Appendix}
The formulation in~\eqref{RNLS-cost-problem} can be interpreted from a Bayesian perspective~\citep{Chen:IFAC:2013}. Bayesian probabilistic estimation is concerned with determining the {\em a posteriori} distribution of $\bm \theta$ given the measurements $\bm y$ or $p( {\bm \theta} | {\bm y} )$, where $p(\cdot | \cdot)$ denotes conditional probability density function (pdf).  According to the Bayes' rule, one has
\begin{align*} 
p(\bm \theta|\bm y) = \frac{p(\bm y|\bm \theta)p(\bm \theta)}{p(\bm y)} \propto  p(\bm y|\bm \theta)p(\bm \theta).
\end{align*}
Consider that the measurement $\bm y$ is corrupted by noise $\bm w$, as often happens in practice. That is,
\begin{align*}
{\bm y} = {\bm \phi} ({\bm \theta}) +{\bm w}.
\end{align*}
If ${\bm w}$ is independent of ${\bm \theta}$ and follows a Gaussian distribution $\mathcal{N} ({\bm 0}, {\bm Q})$, then $p(\bm y|\bm \theta) \sim \mathcal{N} \left( {\bm \phi} ({\bm \theta}) , {\bm Q}\right)$. Meanwhile, $\bm \theta$ is  a Gaussian random vector. Its {\em a priori} pdf before the observation of $\bm y$ is $p(\bm \theta)\sim \mathcal{N}({\bm \theta}_0, {\bm P}_0)$. Then, it follows that 
\begin{align*}
p(\bm \theta|\bm y) \propto  {\rm exp}\left(- \frac {1}{2}\left[\bm y- {\bm \phi} \left(\bm \theta\right) \right]^{\T} \bm Q^{-1}  \left[\bm y- {\bm \phi} \left(\bm \theta\right) \right] \right) {\rm exp}\left( -\frac {1}{2} \left(\bm \theta-\bm \theta_0\right)^{\T} {\bm P_0}^{-1}  \left(\bm \theta-\bm \theta_0\right) \right) .
\end{align*}
From the perspective of estimation, it is desired to find out $\bm \theta$ to maximize $p(\bm \theta|\bm y)$. This is known as  maximum a posteriori (MAP) estimation and can reduce to the problem shown in~\eqref{RNLS-cost-problem} by considering the log likelihood.

\balance
\bibliographystyle{elsarticle-num} 
\bibliography{bibliography}

\end{document}

%% file: headinput.tex
\newcommand{\T}{{\top}}